\documentclass[a4paper,12pt,eqno]{article}
\usepackage{theorem}
\usepackage{latexsym,amssymb,amsfonts,amsmath}
\usepackage{graphicx}
\usepackage{color}

\newtheorem{thm}{Theorem}[section]
\newtheorem{lem}[thm]{Lemma}
\newtheorem{cor}[thm]{Corollary}
\newtheorem{pro}[thm]{Proposition}
\newtheorem{rem}[thm]{Remark}

\theoremheaderfont{\scshape}

\title{{\Large {\bf THE SPECTRA OF THE UNITARY MATRIX OF A 2-TESSELLABLE STAGGERED QUANTUM WALK ON A GRAPH}}
\author{
{\small Norio Konno}\\
{\scriptsize Department of Applied Mathematics, 
Faculty of Engineering, 
Yokohama National University}\\
{\scriptsize Hodogaya, Yokohama 240-8501, Japan}\\
{\scriptsize e-mail: konno@ynu.ac.jp, Tel.: +81-45-339-4205, Fax: +81-45-339-4205}\\
{\small Iwao Sato}\\
{\scriptsize Oyama National College of Technology}\\
{\scriptsize Oyama, Tochigi 323-0806, Japan}\\
{\scriptsize e-mail: isato@oyama-ct.ac.jp. Tel.: +81-285-20-2100, Fax: +81-285-20-2880}\\
{\small Etsuo Segawa} \\
{\scriptsize Graduate School of Information Sciences, Tohoku University} \\
{\scriptsize Sendai 980-8579, Japan}\\}
}
\date{\empty }
\begin{document}
\maketitle

\par\noindent
\begin{small}
\par\noindent
{\bf Abstract}. Recently, the staggered quantum walk (SQW) on a graph is discussed as 
a generalization of coined quantum walks on graphs and Szegedy walks. 
We present a formula for the time evolution matrix of a 2-tessellable SQW on a graph, and so 
directly give its spectra. 
Furthermore, we present a formula for the Szegedy matrix of a bipartite graph by the same method, and so 
give its spectra. 
As an application, we present a formula for the characteristic polynomial of the modified Szegedy matrix 
in the quantum search problem on a graph, and give its spectra.

\footnote[0]{
{\it Abbr. title:} The transition matrix of a quantum walk on a graph
}
\end{small}

\section{Introduction}
As a quantum counterpart of the classical random walk, the quantum walk has recently attracted much attention for various fields. 
The review and book on quantum walks are Ambainis \cite{Ambainis2003}, Kempe \cite{Kempe2003}, Kendon \cite{Kendon2007}, 
Konno \cite{Konno2008b}, Venegas-Andraca \cite{VA}, Manouchehri and Wang \cite{MW}, Portugal \cite{Por 2013},  examples. 

Quantum walks of graphs were studied by many researchers.  
A discrete-time quantum walk on a line was proposed by Aharonov et al \cite{Aha et 1993}. 
In \cite{Aha et 2001}, a discrete-time quantum walk on a regular graph was proposed. 
The Grover walk is a discrete-time quantum walk on a graph which originates from the Grover algorithm. 
The Grover algorithm which was introduced in \cite{Grover} is a quantum search algorithm that performs quadratically faster than 
the best classical search algorithm. 
Using a different quantization procedure, Szegedy \cite{Sze} proposed a new coinless discrete-time quantum walk, 
i.e., the Szegedy walk on a bipartite graph and provided a natural definition of quantum hitting time. 
Also, Szegedy developed quantum walk-based search algorithm, which can detect the presence of a marked vertex 
at a hitting time that is quadratically smaller than the classical average time on ergodic Markov chains. 
Portugal \cite{PSF 2016}, \cite{Por 2016}, \cite{Por 2016II}, defined the staggered quantum walk (SQW) on a graph 
as a generalization of coined quantum walks on graphs and Szegedy walks. 
In \cite{PSF 2016}, \cite{Por 2016}, Portugal studied the relation between SQW and coined quantum walks, Szegedy walks. 
In \cite{Por 2016II}, Portugal presented some properties of 2-tessellable SQW on graphs by using several results of the graph theory. 

Spectra of various quantum walk on a graph were computed by many researchers. 
Related to graph isomorphism problems, Emms et al. \cite{EmmsETAL2006} presented spectra of the Grover matrix
(the time evolution matrix of the Grover walk) on a graph and those of the positive supports of 
the Grover matrix and its square. 
Konno and Sato \cite{KS2011} computed the characteristic polynomials for the Grover matrix and its positive supports of a graph 
by using determinant expressions for several graph zeta functions, and so directly gave their spectra. 
Godsil and Guo \cite{GG2010} gave new proofs of the results of Emms et al. \cite{EmmsETAL2006}.  

In the quantum search problem, the notion of hitting time in classical Markov chains is generalized to quantum hitting time. 
Kempe \cite{Kempe2003II} provided two definitions and proved that a quantum walker hits the opposite corner of an $n$-hypercube in time $O(n)$. 
Krovi and Brun \cite{KB} provided a definition of average hitting time that requires a partial measurement of the position of the walker at each step. 
Kempe and Portugal \cite{KP} discussed the relation between hitting times and the walker's group velocity. 
Szegedy \cite{Sze} gave a definition of quantum hitting time that is a natural generalization of the classical definition of hitting time. 
Magniez et al \cite{MNRS} extended Szegedy's work to non-symmetric ergodic Markov chains. 
Recently, Santos and Portugal \cite{SP} calculated analytically Szegedy's hitting time and 
the probability of finding a set of marked vertices on the complete graph.

The rest of the paper is organized as follows. 
Section 2 states some definitions and notation on graph theory, and gives the definitions of the Grover walk, 
the Szegedy walk, the staggered quantum walk (SQW) on a graph and a short review on the quantum search problem on a graph. 
In Sect. 3, we present a formula for the time evolution matrix of a 2-tessellable SQW on a graph, and so 
give its spectra. 
In Sect. 4, we present a formula for the Szegedy matrix of a bipartite graph, and so 
give its spectra. 
In Sect. 6, we present a formula for the modified time evolution matrix of the duplication of the modified digraph which is 
appeared in the quantum search problem on a graph, and so give its spectra.

\section{Definition of several quantum walks on a graph} 

\subsection{Definitions and notation} 

Graphs treated here are finite. 
Let $G=(V(G),E(G))$ be a connected graph (possibly multiple edges and loops) 
with the set $V=V(G)$ of vertices and the set $E=E(G)$ of unoriented edges $uv$ 
joining two vertices $u$ and $v$. 
Two vertices $u$ and $v$ of $G$ are {\em adjacent} if there exits an edge $e$ joining $u$ and $v$ in $G$. 
Furthermore, two vertices $u$ and $v$ of $G$ are {\em incident} to $e$.  
The {\em degree} $\deg v = \deg {}_G \  v$ of a vertex $v$ of $G$ is the number of edges incident to $v$. 
For a natural number $k$, a graph $G$ is called {\em $k$-regular } if $\deg {}_G \  v=k$ for each vertex $v$ of $G$. 

For $uv \in E(G)$, an arc $(u,v)$ is the oriented edge from $u$ to $v$. 
Set $D(G)= \{ (u,v),(v,u) | uv \in E(G) \} $. 
For $e=(u,v) \in D(G)$, set $u=o(e)$ and $v=t(e)$. 
Furthermore, let $e^{-1}=(v,u)$ be the {\em inverse} of $e=(u,v)$. 
A {\em path} $P=(v_1, v_2 , \ldots , v_{n+1} )$ {\em of length $n$} in $G$ is a sequence of $(n+1)$ vertices such that 
$v_i v_{i+1} \in E(G)$ for $i=1, \ldots ,n$. 
Then $P$ is called a {\em $(v_1 , v_{n+1} )$-path}. 
If $e_i =v_i v_{i+1} (1 \leq i \leq n)$, then we write $P=( e_1 , \ldots e_n )$. 
 
A graph $G$ is called a {\em complete} if any two vertices of $G$ are adjacent. 
We denote the complete graph with $n$ vertices by $K_n $. 
Furthermore, a graph $G$ is called {\em bipartite}, denoted by $G=( V_1 ,V_2 )$ 
if there exists a partition $V(G)= V_1 \cup V_2 $ of $V(G)$ such that 
the vertices in $V_i$ are mutually nonadjacent for $i=1,2$. 
The subsets $V_1 ,V_2 $ of $V(G)$ is called the {\em bipartite set} or the {\em bipartition} of $G$. 
A bipartite graph $G=( V_1 ,V_2 )$ is called {\em complete} if any vertex of $V_1 $ and any vertex of $V_2 $ are 
adjacent. 
If $|V_1 |=m$ and $|V_2 |=n$, then we denote the complete bipartite woth bipartition $V_1 , V_2$ by $K_{m,n} $.

Next, we define two operations of a graph. 
Let $G$ be a connected graph. 
Then a subgraph $H$ of $G$ is called a {\em clique} if $H$ is a complete subgraph of $G$. 
The {\em clique graph} $K(G)$ of $G$ has its vertex set the maximal cliques of $G$, with two vertices 
adjacent whenever they have some vertex of $G$ in common. 
Furthermore, the {\em line graph} $L(G)$ of $G$  has its vertex set the edges of $G$, with two vertices 
adjacent whenever they have some vertex of $G$ in common.

\subsection{The Grover walk on a graph}

A discrete-time quantum walk is a quantum analog of the classical random walk on a graph whose state vector is governed by 
a matrix called the transition matrix.  
Let $G$ be a connected graph with $n$ vertices and $m$ edges, 
$V(G)= \{ v_1 , \ldots , v_n \} $ and $D(G)= \{ e_1 , \ldots , e_m , 
e^{-1}_1 , \ldots , e^{-1}_m \} $. 
Set $d_j = d_{u_j} = \deg v_j $ for $i=1, \ldots ,n$. 
The {\em transition matrix} ${\bf U} ={\bf U} (G)=( U_{ef} )_{e,f \in D(G)} $ 
of $G$ is defined by 
\[
U_{ef} =\left\{
\begin{array}{ll}
2/d_{t(f)} (=2/d_{o(e)} ) & \mbox{if $t(f)=o(e)$ and $f \neq e^{-1} $, } \\
2/d_{t(f)} -1 & \mbox{if $f= e^{-1} $, } \\
0 & \mbox{otherwise.}
\end{array}
\right. 
\]
The matrix ${\bf U} $ is called the {\em Grover matrix} of $G$. 

We introduce the {\em positive support} $\>{\bf F}^+ =( F^+_{ij} )$ of 
a real matrix ${\bf F} =( F_{ij} )$ as follows: 
\[
F^+_{ij} =\left\{
\begin{array}{ll}
1 & \mbox{if $F_{ij} >0$, } \\
0 & \mbox{otherwise.}
\end{array}
\right.
\]

Let $G$ be a connected graph. 
If the degree of each vertex of $G$ is not less than 2, i.e., $\delta (G) \geq 2$, 
then $G$ is called an {\em md2 graph}. 

The transition matrix of a discrete-time quantum walk in a graph 
is closely related to the Ihara zeta function of a graph. 
We stare a relationship between the discrete-time quantum walk and the Ihara zeta function of a graph by Ren et al. \cite{RenETAL}.

Konno and Sato \cite{KS2011} obtained the following formula of the characteristic polynomial of ${\bf U}$ 
by using the determinant expression for the second weighted zeta function of a graph. 

Let $G$ be a connected graph with $n$ vertices and $m$ edges. 
Then the $n \times n$ matrix ${\bf T} (G)=( T_{uv} )_{u,v \in V(G)}$ is given as follows: 
\[
T_{uv} =\left\{
\begin{array}{ll}
1/( \deg {}_G u)  & \mbox{if $(u,v) \in D(G)$, } \\
0 & \mbox{otherwise.}
\end{array}
\right.
\] 
Note that the matrix ${\bf T} (G)$ is the transition matrix of the simple random walk on $G$.

\begin{thm}[Konno and Sato \cite{KS2011}]
\label{thm4.1}
Let $G$ be a connected graph with $n$ vertices $v_1 , \ldots , v_n $ and $m$ edges. Then, for the transition matrix ${\bf U}$ of $G$, we have
\begin{align*}
\det ( \lambda {\bf I}_{2m} - {\bf U} )
&= 
( \lambda {}^2 -1)^{m-n} \det (( \lambda {}^2 +1) {\bf I}_n -2 \lambda {\bf T} (G)) \\
& = \frac{( \lambda {}^2 -1)^{m-n} \det (( \lambda {}^2 +1) {\bf D} -2 \lambda {\bf A} (G))}
{d_{v_1} \cdots d_{v_n }} ,  
\end{align*}
where ${\bf A} (G)$ is the adjacency matrix of $G$, and ${\bf D} =( d_{uv} )$ is the diagonal matrix given by 
$d_{uu} = \deg u \ (u \in V(G))$. 
\end{thm}

From this Theorem, the spectra of the Grover matrix on a graph is obtained by means of those of 
${\bf T} (G)$ (see \cite{EmmsETAL2006}). 
Let $Spec ({\bf F})$ be the spectra of a square matrix ${\bf F}$ .

\begin{cor}[Emms, Hancock, Severini and Wilson \cite{EmmsETAL2006}] 
\label{cor4.2}
Let $G$ be a connected graph with $n$ vertices and $m$ edges. 
The transition matrix ${\bf U}$ has $2n$ eigenvalues of the form 
\[
\lambda = \lambda {}_T \pm i \sqrt{1- \lambda {}^2_T } , 
\]
where $\lambda {}_T $ is an eigenvalue of the matrix ${\bf T} (G)$. 
The remaining $2(m-n)$ eigenvalues of ${\bf U}$ are $\pm 1$ with equal multiplicities. 
\end{cor}

Emms et al. \cite{EmmsETAL2006} determined the spectra of the transition matrix ${\bf U}$ 
by examining the elements of the transition matrix of a graph and 
using the properties of the eigenvector of a matrix.   
And now, we could explicitly obtain the spectra of the transition matrix ${\bf U}$ from 
its characteristic polynomial. 

Next, we state about the positive support of the transition matrix of a graph. 

Emms et al \cite{EmmsETAL2006} expressed the spectra of the positive support ${\bf U}^+ $ of 
the transition matrix of a regular graph $G$ by means of those of the adjacency matrix ${\bf A} (G)$ 
of $G$.

\begin{thm}[Emms, Hancock, Severini and Wilson \cite{EmmsETAL2006}]
\label{thm5.1}
Let $G$ be a connected $k$-regular graph with $n$ vertices and $m$ edges, and 
$\delta (G) \geq 2$. 
The positive support ${\bf U}^+ $ has $2n$ eigenvalues of the form 
\[
\lambda = \frac{\lambda {}_A }{2} \pm i \sqrt{k-1- \lambda {}^2_A /4} , 
\]
where $\lambda {}_A $ is an eigenvalue of the matrix ${\bf A} (G)$. 
The remaining $2(m-n)$ eigenvalues of ${\bf U}^+$ are $\pm 1$ with equal multiplicities. 
\end{thm}

Godsil and Guo \cite{GG2010} presented a new proof of Theorem \label{thm5.1} 
by using linear algebraic technique. 

Konno and Sato \cite{KS2011} obtained the following formula of 
the characteristic polynomial of ${\bf U}^+$ by using the determinant expression for the Ihara zeta function 
of a graph, and directly presented the spectra of the positive support ${\bf U}^+ $ of 
the transition matrix of a regular graph $G$.

\subsection{The Szegedy quantum walk on a bipartite graph} 

Let $G=(X \sqcup Y,E)$ be a connected bipartite graph with partite set $X$ and $Y$. 
Moreover, set $|V(G)|= \nu $, $|E|=|E(G)|= \epsilon $, $|X|=m$ and $|Y|=n$. 
Then we consider the Hilbert space ${\cal H} = \ell {}^2 (E)=span \{  |e \rangle \mid e \in E \}$.
Let $p: E \rightarrow [0,1]$ and $q: E \rightarrow [0,1]$ be the functions such that 
\[
\sum_{X(e)=x} p(e)= \sum_{Y(e)=y} q(e)=1, \forall x\in X , \forall y \in Y , 
\]
where $X(e)$ and $Y(e)$ are the vertex of $e$ belonging to $X$ and $Y$, respectively. 

For each $x \in X$ and $y \in Y$, let 
\[
| \phi {}_x \rangle = \sum_{X(e)=x} \sqrt{p(e)} |e \rangle \  and \ 
| \psi {}_y \rangle = \sum_{Y(e)=y} \sqrt{q(e)} |e \rangle . 
\] 
From these vectors, we construct two $\epsilon \times \epsilon $ matrices ${\bf R}_0 $ and ${\bf R}_1 $ 
as follows: 
\[
{\bf R}_0  =2 \sum_{x \in X} | \phi {}_x \rangle \langle \phi {}_x | - {\bf I}_{ \epsilon }, \ 
{\bf R}_1  =2 \sum_{y \in Y} | \psi {}_y \rangle \langle \psi {}_y | - {\bf I}_{ \epsilon } 
\]
Furthermore, we define an $ \epsilon \times \epsilon $ matrix ${\bf W} $ as follows: 
\[
{\bf W} = {\bf R}_1 {\bf R}_0 . 
\]
Note that two matrices ${\bf R}_0 $ and ${\bf R}_1 $ are unitary, and ${\bf R}^2_0 ={\bf R}^2_1 = {\bf I}_{mn} $. 

The quantum walk on $G$ with ${\bf W} $ as a time evolution matrix is called 
{\em Szegedy walk} on $G$, and the matrix ${\bf W} $ is called the {\em Szegedy matrix} 
of $G$.

\subsection{The staggered quantum walk on a graph} 

Let $G$ be a connected graph with $ \nu $ vertices and $ \epsilon $ edges. 
Furthermore, let ${\cal H}^{\nu } $ be the Hilbert space generated by the vertices of $G$. 
We take a standard basis as $\{|u\rangle \;\mid\; u\in V \}$. 
In general, a unitary and Hermitian operator ${\bf V} $ on ${\cal H}^{\nu } $ can be written by 
\[
{\bf V} = \sum_x | \psi {}^+_x \rangle \langle \psi {}^+_x | - \sum_y | \psi {}^-_y \rangle \langle \psi {}^-_y | , 
\]
where the set of vectors $| \psi {}^+_x \rangle$ is a normal orthogonal basis of $(+1)$-eigenspace, and 
the set of vectors $| \psi {}^-_x \rangle$  is a normal orthogonal basis of $(-1)$-eigenspace. 
Since 
\[
\sum_x | \psi {}^+_x \rangle \langle \psi {}^+_x | + \sum_y | \psi {}^-_y \rangle \langle \psi {}^-_y |= {\bf I} , 
\]
we obtain 
\[
{\bf U} =2 \sum_x | \psi {}^+_x \rangle \langle \psi {}^+_x | - {\bf I} \cdots (*) . 
\]

A unitary and Hermitian matrix ${\bf V} $ in ${\cal H}^{\nu } $ given by (*) is called an 
{\em orthogonal reflection} of $G$ if the set of the orthogonal set of $(+1)$-eigenvectors $\{| \psi {}^+_x \rangle\}_x $ obeying the following properties: 
\begin{enumerate} 
\item If the $i$-th entry of $| \psi {}^+_x \rangle $ for a fixed $x$ is nonzero, the $i$-th entry of 
the other $(+1)$-eigenvectors are zero, that is, if $\langle i|\psi_x \rangle \neq 0$, then $\langle i|\psi_{x'} \rangle =0$ for any $x'\neq x$; 
\item The vector $\sum_x | \psi {}^+_x \rangle $ has no zero entries. 
\end{enumerate} 

Next, a {\em polygon} of a graph $G$ induced by a vector $| \psi \rangle \in {\cal H}^{\nu} $ is a clique.
That is, two vertices of $G$ are adjacent if the corresponding entries of $| \psi \rangle $ in the basis associated with 
$G$ are nonzero. Thus if $\langle u|\psi\rangle \neq 0$ and $\langle v|\psi\rangle \neq 0$, then $u$ is connected to $v$. 
A vertex belongs to the polygon if and only if it corresponding entry in $| \psi \rangle $ is nonzero. 
An edge belongs to the polygon if and only if the polygon contains the endpoints of the edge. 

A {\em tessellation} induced by an orthogonal reflection ${\bf V} $ of $G$ is the union of the polygons induced by the 
$(+1)$-eigenvectors $\{|\psi {}^+_x \rangle\}_x $ of ${\bf V} $ described in the above. 
The {\em staggered quantum walk(SQW)} on $G$ associated with the Hilbert space ${\cal H}^{\nu } $ is driven by 
\[
{\bf U} = {\bf U}_1 {\bf U}_0 ,
\]
where ${\bf U}_0 $ and ${\bf U}_1 $ are orthogonal reflections of $G$. 
The union of the tessellations $\alpha $ and $\beta $ by ${\bf U}_0 $ and ${\bf U}_1 $ must cover the edges of $G$. 
Furthermore, set $\alpha = \{ \alpha {}_1 , \ldots , \alpha {}_m \} $ and $ \beta = \{ \beta {}_1 , \ldots , \beta {}_n \} $. 
Then ${\bf U}_0 $ and ${\bf U}_1 $ are given as follows:  
\[
{\bf U}_0 =2 \sum^m_{k=1} | \alpha {}_k \rangle \langle \alpha {}_k | - {\bf I}_{ \nu} , \ 
{\bf U}_1 =2 \sum^n_{l=1} | \beta {}_l \rangle \langle \beta {}_l | - {\bf I}_{ \nu} ,  
\]
where 
\[
| \alpha {}_k \rangle = \sum_{ k^{\prime } \in \alpha {}_k } a_{k k^{\prime } } | k^{\prime } \rangle \  (1 \leq k \leq m), \  
| \beta {}_l \rangle = \sum_{ l^{\prime } \in \beta {}_l } b_{l l^{\prime } } | l^{\prime } \rangle \  (1 \leq l \leq n) . 
\]

A graph $G$ is {\em 2-tessellable} if the following conditions holds: 
\[
V( \alpha {}_1 ) \sqcup \cdots \sqcup V( \alpha {}_m )= V( \beta {}_1 ) \sqcup \cdots \sqcup V( \beta {}_n )=V(G) 
\]
and 
\[   
( E( \alpha {}_1 ) \sqcup \cdots \sqcup E( \alpha {}_m )) \cup ( E( \beta {}_1 ) \sqcup \cdots \sqcup E( \beta {}_n ))=E(G) ,  
\] 
where ${\bf U} = {\bf U}_1 {\bf U}_0 $ is the unitary matrix of a SQW on $G$, and 
$\alpha = \{ \alpha {}_1 , \ldots , \alpha {}_m \} $ and $ \beta = \{ \beta {}_1 , \ldots , \beta {}_n \} $ are 
tessellations of ${\bf U} $ corresponding to ${\bf U}_0 $ and ${\bf U}_1 $, respectively.

\subsection{The quantum search problem on a graph} 

Let $G=(V,E)$ be a connected non-bipartite graph with $n$ vertices and $\epsilon$ edges which may have multiple edges and self loops.  
Let $E_H(u,v)$ be the subset of the edge set of a graph $H$ connecting between vertices $u$ and $v$.  
It holds $\sqcup_{u,v,\in V(H)} E_H(u,v)=E(H)$, where ``$\sqcup$" means the disjoint union. 
%
We want to set the quantum search of an element of $M\subset V$ by the Szegedy walk.
The Szegedy walk is defined by a bipartite graph. 
To this end, we construct the duplication of $G$. 
The {\em duplication} $G_2 $ of $G$ is defined as follows: 
The duplication graph $G_2$ of $G$ is defined as follows. 
\begin{align*}
V(G_2)=V \sqcup V', 
\end{align*} 
where $v'$ is the copy of $v\in V$, therefore $V'=\{v': v\in V\}$. 
The edge set of $E(G_2)$ is denoted by
\begin{equation*}
E_G(u,v)\subset E(G) \Leftrightarrow E_{G_2}(u,v') \subset E(G_2)
\end{equation*}
The end vertex of $e\in E(G_2)$ included in $V$ is denoted by $V(e)$, and one included in $V'$ is denoted by $V'(e)$. 
We consider two functions $p: E(G_2) \rightarrow [0,1]$ and $q: E(G_2) \rightarrow [0,1]$ be the functions such that 
\[
\{p(e) \;|\; e\in E_{G_2}(u,v')\} = \{q(e) \;|\; e\in E_{G_2}(u',v)\}
\]
where 
\[
\sum_{V(e)=x} p(e)= \sum_{V'(e)=y} q(e)=1, \forall x\in V , \forall y \in V' .  
\]
The $2n \times 2n$ stochastic matrix ${\bf P}$ is denoted by 
	\[ ({\bf P})_{u,v}=p_{uv}
        	= \begin{cases} \sum_{V(e)=u,\;V'(e)=v} p(e) & \text{if $u\in V$, $v\in V'$, } \\ 
                		\sum_{V'(e)=u,\;V(e)=v} q(e) & \text{if $u\in V'$, $v\in V$, } \\
                                0 & \text{otherwise.}
                \end{cases} \]
Let $V(G)= \{ v_1 , \ldots , v_n \} $ and $M= \{ v_{n-m+1} , \ldots , v_n \} $. 
Then define the {\em modified digraph} $ \vec{G} $ from $G$ as follows: 
The {\em modified digraph} $ \vec{G} $ with respect to $M$ is obtained from the symmetric digraph $D_G$ 
by converting all arcs leaving from the marked vertices into loops. 
In the duplication $G_2$, the set $M_2$ of marked vertices is 
\[
M_2 =M \cup \{ u^{\prime } \mid u \in M \} . 
\] 
The modified bipartite digraph $\vec{G}_2 $ is obtained from the symmetric digraph of $G_2$ by deleting all arcs leaving 
from the marked vertices of $G_2 $, but keeping the incoming arcs to the marked vertices of $G_2 $ and all other arcs 
unchanged. 
Moreover, we add new $2m=2|M|$ arcs $(u, u^{\prime } ), (u^{\prime } , u)$ for $u \in M$. 
Then the modified bipartite digraph $\vec{G}_2 $ is obtained by taking the duplication of $\vec{G}$. 
More precisely, let $A(G_2)=D(G_2 )$ be the set of symmetric arcs naturally induced by $E(G_2)$, then 
\begin{align*}
V(\overrightarrow{G}_2) &= V(G_2), \\
A(\overrightarrow{G}_2) &= \{a\in A(G_2) \;|\; o(a)\notin M\cup M'\} \cup \{ a,a^{-1} \;|\; o(a)= u,t(a)=u' ,\;u\in M   \}.
\end{align*}
Here $M'\subset V'$ is the copy of $M$. We put the first arcset in RHS by $A_2$, and the second one by $N_2$. 
The modified graph $\vec{G}_2 $ keeps the bipartiteness with $V$ and $V'$.  
Thus once a random walker steps in $M_2$, then she will be trapped in $M_2$ forever. 

We want to induce the Szegedy walk from this absorption picture into $M_2$ of $\vec{G}_2$. 
The Szegedy is denoted by non-directed edges of the bipartite graph. 
So we consider the support of $A(\overrightarrow{G}_2)$ by $E_2:=[A(\overrightarrow{G}_2)]:= \{[a] \;|\; a\in A(\overrightarrow{G}_2) \}$. 
Here $[a]$ is the edge obtained by removing the direction of the arc $a$. 
Thus $E_2=[A_2]\sqcup [N_2]$, and remark that $[N_2]$ describes the set of the matching between $m$ and $m'$ for $m\in M$. 
Taking the following modification to $p$ and $q$, the above absorption picture of a classical walk is preserved by the following random walk 
as follows. For $e\in E_2$, 
	\begin{align*}
        p'(e) &=
        \begin{cases}
        p(e) & \text{if $V(e)\notin M$,} \\
        1 & \text{if $e\in [N_2]$,} \\
        0 & \text{if $V(e)\in M$ and $V'(e)\notin M'$,} 
        \end{cases}
        \\
        q'(e) &=
        \begin{cases}
        q(e) & \text{if $V'(e)\notin M$,} \\
        1 & \text{if $e\in [N_2]$.} \\
        0 & \text{if $V'(e)\in M'$ and $V(e)\notin M$,}
        \end{cases}
        \end{align*} 
The modified $2n \times 2n$ stochastic matrix ${\bf P}'$ is given by changing $p$ and $q$ to $p'$ and $q'$ as follows: 
		\[ ({\bf P}^{\prime } )_{u,v}=p'_{uv}
        	= \begin{cases} \sum_{V(e)=u,\;V'(e)=v} p'(e) & \text{if $u\in V$, $v\in V'$, } \\ 
                		\sum_{V'(e)=u,\;V(e)=v} q'(e) & \text{if $u\in V'$, $v\in V$, } \\
                                0 & \text{otherwise.}
                \end{cases} \]

If there exists a marked element connecting to another marked element in $G$, then 
such an edge is omitted by the procedure of the deformation to $\vec{G}_2$, thus $[A_2] \subset E(G_2)$, 
on the other hand, otherwise, $[A_2]=E(G_2)$. 
We set $F_2=\{ e\in E(G_2) \;|\; V(e),V'(e)\in M_2 \}$. 
Now we are considering a quantum search setting without any connected information about marked elements, 
so we want to set the initial state as a usual way,  
	\[ \psi_0= \sum_{e\in E( G_2 )} \sqrt{p(e)}|e\rangle. \]
However in the above situation, that is, $F_2\neq \emptyset$, since an original edge of $G_2$ is omitted, we cannot define this initial state. 
So we expand the considering edge set 
	\[ E_M:= E_2 \cup F_2. \]
We re-define $p'$ and $q'$ whose domain is changed to $E_M$: 
for every $e\in E_M$. 
	\begin{align*}
        p'(e) &=
        \begin{cases}
        p(e) & \text{if $V(e)\notin M$,} \\
        1 & \text{if $e\in [N_2]$,} \\
        0 & \text{otherwise} 
        \end{cases}
        \\
        q'(e) &=
        \begin{cases}
        q(e) & \text{if $V'(e)\notin M$,} \\
        1 & \text{if $e\in [N_2]$,} \\
        0 & \text{otherwise,}
        \end{cases}
        \end{align*}  
Remark that the above ``otherwise" in the definition of $p^{\prime } $ is equivalent to the situation of ``$V(e)\in M$ and $V'(e)\notin M'$" or ``$e\in F_2$". 

Now we are ready to give the setting of quantum search problem. 
Remark that $E_M=2\epsilon+m$. 
For each $x \in V$ and $y \in V'$, let 
\[
| \phi {}^{\prime }_x \rangle = \sum_{V(e)=x} \sqrt{p^{\prime } (e)} |e\rangle \  and \ 
| \psi {}^{\prime }_y \rangle = \sum_{V'(f)=y} \sqrt{q^{\prime } (f)} |f\rangle . 
\] 
From these unit vectors, we construct two $(2\epsilon+m) \times (2\epsilon+m)$ matrices ${\bf R}^{\prime }_0 $ 
and ${\bf R}^{\prime }_1 $ as follows: 
\[
{\bf R}^{\prime }_0  =2 \sum_{x \in V} | \phi {}^{\prime }_x \rangle \langle \phi {}^{\prime }_x | - {\bf I}_{2\epsilon+m }, \ 
{\bf R}^{\prime }_1  =2 \sum_{y \in V'} | \psi {}^{\prime }_y \rangle \langle \psi {}^{\prime }_y | - {\bf I}_{2\epsilon+m } 
\]
Furthermore, we define an $(2\epsilon+m)\times (2\epsilon+m) $ matrix ${\bf W}^{\prime } $ as follows: 
\[
{\bf W}^{\prime } = {\bf R}^{\prime }_1 {\bf R}^{\prime }_0 . 
\]
Then ${\bf W}^{\prime } $ is the time evolution matrix of the modified Szegedy walk on $\ell^2(E_M)$.

The initial condition of the quantum walk is 
\[
| \psi (0) \rangle = \frac{1}{ \sqrt{n}} \sum_{e \in E( G_2 )} \sqrt{p(e)} |e \rangle .
\]
Note that $| \psi (0) \rangle $ is defined using a random walk on $G$ determined by $p$, and it is invariant under the action of 
${\bf W} ={\bf R}_1 {\bf R}_0 $ associated with $G$(see \cite{Por 2013}). 
We assume that $p_{uv'} =p_{v'u} , u,v \in V(G)$ for the stochastic matrix ${\bf P}$. 
Then ${\bf P}$ is doubly stochastic. 
Let 
\[
|\psi (t) \rangle =( {\bf W}^{\prime } )^t | \psi (0) \rangle  
\]
and 
\[
F(T)= \frac{1}{T+1} \sum^T_{t=0}||| \psi(t)\rangle-|\psi(0)\rangle||^2. 
\]
Then the {\em quantum hitting time} $H_{P,M} $ of a quantum walk on $G$ is fined as the smallest number of steps $T$ such that 
\[
F(T) \geq 1- \frac{m}{n} , 
\]
where $n=|V(G)|$ and $m=|M|$. 
The quantum hitting time is evaluated by the square of the spectral gap of the $n\times n$ matrix ${\bf P}_M$: 
	\[({\bf P}_M)_{u,v}=\begin{cases} p_{u,v'} & \text{ if $u,v\notin M$, }\\ 0 & \text{otherwise. } \end{cases}  \]

\section{Key method}
From now on, we will attempt to three cases of the characteristic polynomials of the time evolution; ``a 2-tessellable staggered quantum matrix", 
``Szegedy matrix " and ``modified Szegedy matrix of quantum search". 
To this end, we provide the key lemma. 
\begin{thm}\label{key}
Let ${\bf A}$ and ${\bf B}$ be $N \times s$ and $N \times t$ complex valued isometry matrices, that is, 
	\[ {}^*{\bf A} {\bf A}= {\bf I}_s, \mathrm{\;and\;} ^*{\bf B} {\bf B}={\bf I}_t,  \] 
where ${}^*{\bf Y}$ is the conjugate and transpose of ${\bf Y}$. 
Putting ${\bf U}={\bf U}_B{\bf U}_A$ with 
${\bf U}_B=(2{\bf B} \ {}^*{\bf B}-{\bf I}_N)$ and ${\bf U}_A=(2{\bf A} \ {}^*{\bf A}-{\bf I}_N)$, 
we have
 	\begin{align*}  
        \det ({\bf I}_N-u{\bf U} )  
        &=(1-u)^{N-(s+t)} (1+u)^{s-t} \det\left[ (1+u)^2 {\bf I}_t -4u {}^*{\bf B} {\bf A} {}^*{\bf A}{\bf B} \right], \\
        &=(1-u)^{N-(s+t)} (1+u)^{t-s} \det\left[ (1+u)^2 {\bf I}_s -4u {}^*{\bf A} {\bf B} {}^*{\bf B}{\bf A} \right].
        \end{align*}
\end{thm}
\noindent{ \bf Proof}: 
At first, we have 
	\[ \det ({\bf I}_N-u{\bf U} )=\det ({\bf I}_N-u{\bf U}_B{\bf U}_A ) . \]
Therefore once we can show the first equality, then changing the variables by $A\leftrightarrow B$ and $t\leftrightarrow s$, we have the second equality. 

Now we will show the first equality. 
\[
\begin{array}{rcl}
\  &   & \det ( {\bf I}_{ N } -u {\bf U} )= \det ( {\bf I}_{ N } -u {\bf U}_B {\bf U}_A ) \\ 
\  &   &                \\ 
\  & = & 
\det ({\bf I}_{N } -u(2 {\bf B} \ {}^*{\bf B} - {\bf I}_{ N } )
(2 {\bf A} \ {}^t {\bf A} - {\bf I}_{ N } )) \\ 
\  &   &                \\ 
\  & = & 
\det ({\bf I}_{N } -2u {\bf B} \ {}^* {\bf B} (2 {\bf A} \ {}^* {\bf A} - {\bf I}_{ N } )
+u(2 {\bf A} \ {}^* {\bf A} - {\bf I}_{ N } )) \\ 
\  &   &                \\ 
\  & = & 
\det ((1-u) {\bf I}_{N } +2u {\bf A} \ {}^* {\bf A} -2u {\bf B} \ {}^* {\bf B} (2 {\bf A} \ {}^* {\bf A} - {\bf I}_{ N } )) \\ 
\  &   &                \\ 
\  & = & (1-u )^{N } 
\det ( {\bf I}_{N } + \frac{2u}{1-u} {\bf A} \ {}^* {\bf A} - \frac{2u}{1-u} {\bf B} \ {}^* {\bf B} (2 {\bf A} \ {}^* {\bf A} - {\bf I}_{ N } )) \\ 
\  &   &                \\ 
\  & = & (1-u )^{N } 
\det ( {\bf I}_{N } - \frac{2u}{1-u} {\bf B} \ {}^* {\bf B} (2 {\bf A} \ {}^* {\bf A} - {\bf I}_{ N } )
( {\bf I}_{N } + \frac{2u}{1-u} {\bf A} \ {}^* {\bf A})^{-1} ) \det ( {\bf I}_{N } + \frac{2u}{1-u} {\bf A} \ {}^* {\bf A} ) . 
\end{array}
\]

If ${\bf A}'$ and ${\bf B}'$are a $m \times n $ and $n \times m$ 
matrices, respectively, then we have 
\[
\det ( {\bf I}_{m} - {\bf A}' {\bf B}' )= 
\det ( {\bf I}_n - {\bf B}' {\bf A}' ) . 
\] 

Thus, we have 
\[
\begin{array}{rcl}
\det ( {\bf I}_{N } + \frac{2u}{1-u} {\bf A} \ {}^* {\bf A} ) 
& = & \det ( {\bf I}_{n} + \frac{2u}{1-u} \  {}^* {\bf A} {\bf A} ) \\
\  &   &                \\ 
\  & = & \det ( {\bf I}_{s} + \frac{2u}{1-u} {\bf I}_n ) \\
\  &   &                \\ 
\  & = & (1+ \frac{2u}{1-u} )^s = \frac{(1+u)^s }{(1-u)^s } . 
\end{array}
\] 
Furthermore, we have 
\[
\begin{array}{rcl}
\  &   & ( {\bf I}_{N } + \frac{2u}{1-u} {\bf A} \ {}^* {\bf A} )^{-1} \\ 
\  &   &                \\ 
\  & = & {\bf I}_{N } - \frac{2u}{1-u} {\bf A} \ {}^* {\bf A} +( \frac{2u}{1-u} )^2 {\bf A} \ {}^* {\bf A} {\bf A} \ {}^* {\bf A} 
- ( \frac{2u}{1-u} )^3 {\bf A} \ {}^* {\bf A} {\bf A} \ {}^* {\bf A} {\bf A} \ {}^* {\bf A} + \cdots \\
\  &   &                \\ 
\  & = & {\bf I}_{N } - \frac{2u}{1-u} {\bf A} \ {}^* {\bf A} +( \frac{2u}{1-u} )^2 {\bf A} \ {}^* {\bf A}
- ( \frac{2u}{1-u} )^3 {\bf A} \ {}^* {\bf A} + \cdots \\
\  &   &                \\ 
\  & = & {\bf I}_{N } - \frac{2u}{1-u} (1 - \frac{2u}{1-u} +( \frac{2u}{1-u} )^2 - \cdots ) {\bf A} \ {}^* {\bf A}  \\
\  &   &                \\ 
\  & = & {\bf I}_{N } - \frac{2u}{1-u} /(1+ \frac{2u}{1-u} ) {\bf A} \ {}^* {\bf A} 
={\bf I}_{N } - \frac{2u}{1+u} {\bf A} \ {}^* {\bf A} .
\end{array}
\]

Therefore, it follows that 
\[
\begin{array}{rcl}
\  &   & \det ( {\bf I}_{N } -u {\bf U} ) \\ 
\  &   &                \\ 
\  & = & (1-u )^{N } 
\det ( {\bf I}_{N } - \frac{2u}{1-u} {\bf B} \ {}^* {\bf B} (2 {\bf A} \ {}^* {\bf A} - {\bf I}_{ N } )
( {\bf I}_{N } - \frac{2u}{1+u} {\bf A} \ {}^* {\bf A}))  \frac{(1+u)^s }{(1-u)^s } \\ 
\  &   &                \\ 
\  & = & (1-u )^{N -s} (1+u)^s 
\det ( {\bf I}_{N } + \frac{2u}{1-u} {\bf B} \ {}^* {\bf B} ( {\bf I}_{ N } - \frac{2}{1+u} {\bf A} \ {}^* {\bf A} )) \\ 
\  &   &                \\ 
\  & = & (1-u )^{N -s} (1+u)^s 
\det ( {\bf I}_{t} + \frac{2u}{1-u} \ {}^* {\bf B} ( {\bf I}_{ N } - \frac{2}{1+u} {\bf A} \ {}^* {\bf A} ) {\bf B} ) \\ 
\  &   &                \\ 
\  & = & (1-u )^{N -s} (1+u)^s 
\det ( {\bf I}_{t} + \frac{2u}{1-u} \ {}^* {\bf B} {\bf B} - \frac{4u}{1-u^2 } \ {}^* {\bf B} {\bf A} \ {}^* {\bf A} {\bf B} ) \\ 
\  &   &                \\ 
\  & = & (1-u )^{N -s} (1+u)^s 
\det ( {\bf I}_{t} + \frac{2u}{1-u} {\bf I}_t - \frac{4u}{1-u^2 } \ {}^* {\bf B} {\bf A} \ {}^* {\bf A} {\bf B} ) \\ 
\  &   &                \\ 
\  & = & (1-u )^{N -s} (1+u)^s 
\det ( \frac{1+u}{1-u} {\bf I}_{t}- \frac{4u}{1-u^2 } \ {}^* {\bf B} {\bf A} \ {}^* {\bf A} {\bf B} ) \\ 
\  &   &                \\ 
\  & = & (1-u )^{N - s-t } (1+u)^{s-t}  
\det ((1+u )^2 {\bf I}_{t}-4u \ {}^* {\bf B} {\bf A} \ {}^* {\bf A} {\bf B} ) . 
\end{array}
\]
$\Box$

%
We put ${\bf T}_{BA}:={}^*{\bf B}{\bf A}$ and ${\bf T}_{AB}:={}^*{\bf A}{\bf B}$. Thus ${}^*{\bf T}_{BA}={\bf T}_{AB}$. 

\begin{lem} 
For any eigenvalue $ \lambda_{q} $ of ${\bf T}_{BA}{\bf T}_{AB}$, 
\[
0\leq  \lambda_q \leq 1 . 
\]
\end{lem}

{\em Proof }.  At first, let  
\[
{\bf T}_{BA}{\bf T}_{AB} f= \lambda_{q} f .
\]
Then we have 
\[
\begin{array}{rcl}
| \lambda_q |^2 ||f||^2 & = & || {}^* {{\bf B}} {\bf A} {}^* \overline{{\bf A}} {\bf B} f||^2 \\
\  &   &                \\ 
\  & = & \langle {}^* {{\bf A}} {\bf B} f, {}^* {{\bf A}} {\bf B} f \rangle \\
\  &   &                \\ 
\  & = & \langle {}^t {{\bf B}} f, {\bf A} {}^* {{\bf A}} {\bf B} f \rangle \\
\  &   &                \\ 
\  & \leq & \langle {\bf B} f, {\bf B} f \rangle \\
\  &   &                \\ 
\  & = & \langle f, {}^* {{\bf B}} {\bf B} f \rangle \\
\  &   &                \\ 
\  & = & \langle f, f \rangle =||f|| . 
\end{array}
\]
Thus, 
\[
| \lambda_q | \leq 1 . 
\]
Since $\langle g, {\bf T}_{BA}{\bf T}_{AB}g\rangle \geq 0$ for every $g$, we have $0\leq =\lambda_{q} $ holds. Therefore $\mu\in [0,1]$
$\Box$ 
\begin{rem} Let $s \geq t$. 
Then it holds 
 \[ Spec({\bf T}_{AB}{\bf T}_{BA})=\{0\}^{s-t} \cup Spec({\bf T}_{BA}{\bf T}_{AB}) , \] 
where $\{0\}^{s-t} $ is the multi-set of $s-t$ 0. 
Thus $0\leq \lambda_p \leq 1$ for any $\lambda_p\in Spec(T_{AB}T_{BA})$. 
\end{rem}
\begin{cor}\label{cor1}
For the unitary matrix ${\bf U} = {\bf U}_B {\bf U}_A $, we have    
\[
\det ( \lambda  {\bf I}_{ N } - {\bf U} )=( \lambda -1 )^{N-s-t} ( \lambda +1 )^{s-t} \det (( \lambda +1)^2 {\bf I}_t -4 \lambda {\bf T}_{BA}{\bf T}_{AB} ). 
\]
\end{cor}
{\em Proof }.  Let $u=1/ \lambda $. 
Then, by Theorem 3.1, we have 
\[
\det ({\bf I}_{ N } -1/  \lambda   {\bf U} )=(1-1/ \lambda )^{ N-s-t} (1+1/ \lambda )^{s-t} \det ((1+1/ \lambda )^2 {\bf I}_t -4/\lambda {\bf T}_{BA}{\bf T}_{AB} ) ,  
\]
and so, 
\[
\det ( \lambda  {\bf I}_{ N } - {\bf U} )=( \lambda -1 )^{N-s-t} ( \lambda +1 )^{s-t} \det (( \lambda +1)^2 {\bf I}_t -4 \lambda {\bf T}_{BA}{\bf T}_{AB} ) . 
\]
$\Box$
\begin{cor}\label{cor2q}
Set 
$Spec({\bf T}_{BA}{\bf T}_{AB})=\{\lambda_{q,1},\dots,\lambda_{q,t}\}$
with $0 \leq \lambda_{q,1} \leq \cdots \leq \lambda_{q,t} \leq 1$. 
Moreover the two solutions of 
	\[ \lambda^2-2(2\lambda_{q,j}-1)\lambda+1=0 \]
is denoted by $\alpha_j^{(\pm)}$. 
Then $N$ eigenvalues of ${\bf U}$ are described as follows: 
\begin{enumerate}
\item $|N-(s+t) |$-multiple eigenvalue: $1$; 
\item $|t-s|$-multiple eigenvalue: $(-1)$;
\item $2(\mathrm{Min}\{t, N-s \} - \mathrm{Max}\{ 0,t-s \} )$ eigenvalues:
   \[\alpha_j^{(\pm)}, \;(j=\mathrm{Max}\{1,t-s+1\},\cdots,\mathrm{Min}\{t,N-s\}). \]
\end{enumerate}
Here an expression for $\alpha_j^{(\pm)}$ is 
	\[ \alpha_j^{(\pm)}=e^{\pm 2 \sqrt{-1} \arccos \sqrt{\lambda_{q,j}} }. \]
\end{cor}
\begin{rem}
It holds 
	\[ |N-(s+t)|+|t-s|+2(\mathrm{Min}\{t, N-s \} - \mathrm{Max}\{ 0,t-s \} )=N. \]
In particular, 
\begin{enumerate}
\item If $t < s$, then $\lambda_{q,1} =\cdots= \lambda_{q,s-t} = 0$. 
\item If $N < t+s$, then $\lambda_{q,N-s+1} =\cdots= \lambda_{q,t} = 1$. 
\end{enumerate}
\end{rem}
\begin{cor}\label{cor2p}
Set 
$Spec({\bf T}_{AB}{\bf T}_{BA})=\{\lambda_{p,1},\dots,\lambda_{p,s}\}$
with $0 \leq \lambda_{p,1} \leq \cdots \leq \lambda_{p,s} \leq 1$. 
Moreover the two solutions of 
	\[ \lambda^2-2(2\lambda_{p,j}-1)\lambda+1=0 \]
is denoted by $\beta_j^{(\pm)}$. 
Then $N$ eigenvalues of ${\bf U}$ are described as follows: 
\begin{enumerate}
\item $|N-(s+t) |$-multiple eigenvalue: $1$; 
\item $|t-s|$-multiple eigenvalue: $(-1)$;
\item $2(\mathrm{Min}\{s, N-t \} - \mathrm{Max}\{ 0,s-t \} )$ eigenvalues:
   \[\beta_j^{(\pm)}, \;(j=\mathrm{Max}\{1,s-t+1\},\cdots,\mathrm{Min}\{s,N-t\}). \]
Here an expression for $\beta_j^{(\pm)}$ is 
	\[ \beta_j^{(\pm)}=e^{\pm 2 \sqrt{-1} \arccos \sqrt{\lambda_{p,j}} }. \]
\end{enumerate}
\end{cor}
\begin{rem}\noindent 
\begin{enumerate}
\item If $s < t$, then $\lambda_{p,1} =\cdots= \lambda_{p,t-s} = 0$. 
\item If $N < t+s$, then $\lambda_{p,N-t+1} =\cdots= \lambda_{p,s} = 1$. 
\end{enumerate}
\end{rem}

Once we show Corollary \ref{cor2q}, then Corollary \ref{cor2p} automatically holds by Theorem 3.1. 
So in the following we give a proof of Corollary \ref{cor2q}. \\ 
\\ \noindent 

\noindent {\bf Proof of Corollary \ref{cor2q}}: \\
By Corollary 3.4, we can rewrite the characteristic polynomial of ${\bf U}$ by 
\[
\begin{array}{rcl}
\  &   & \det ( \lambda  {\bf I}_{N } - {\bf U} ) \\
\  &   &                \\ 
\  & = & ( \lambda -1 )^{N - (s+t) } ( \lambda +1 )^{s-t} 
\prod_{j=1}^{t} (( \lambda +1)^2 -4 \lambda {}_{q,j} \lambda ) \\  
\  &   &                \\ 
\  &   & =( \lambda -1 )^{N - (s+t) } ( \lambda +1 )^{s-t} 
\prod_{ j=1}^t ( \lambda {}^2 -2(2 \lambda {}_{q,j} -1) \lambda +1) . 
\end{array}
\]
We put the two solution of $ \lambda {}^2 -2(2 \lambda {}_{q,j} -1) \lambda +1=0$ by $\alpha_{j}^{(\pm)}$. 
Then 
\[ \det ( \lambda  {\bf I}_{N } - {\bf U} )
	=( \lambda -1 )^{N - (s+t) } ( \lambda +1 )^{s-t} \prod_{ j=1}^t ( \lambda-\alpha_j^{(+)})( \lambda-\alpha_j^{(-)})  \]
Concerning that RHS is an $N$-th degree polynomial of $\lambda$, we consider the four cases with respect to the signes of 
$N-(s+t)$ and $s-t$. 
\begin{enumerate}
\item $N-(s+t)\geq 0$, $s-t\geq 0$ case: \\
we directly obtain 
$(N-s-t)$-multiple eigenvalue $1$, 
$(s-t)$-multiple eigenvalue $-1$ and 
$2t$ eigenvalues $\alpha_{q,j}^{(\pm)}$ $(j=1,\dots,t)$.
\item $N-(s+t)\geq 0$, $s-t<0$ case: \\
Since $s-t<0$, $(\lambda+1)^{s-t}$ is a negative power term. To cancel down it, 
$\{ ( \lambda-\alpha_j^{(+)}),\;( \lambda-\alpha_j^{(-)}) \}_{j=1}^t$ must contain $(t-s)$ terms of $(\lambda+1)$. 
Remark that if $\lambda=-1$, then $\lambda_{q,j}=0$ from the above quadratic equation. So $\lambda_{q,1}=\cdots=\lambda_{q,t-s}=0$.
By the above consideration, the characteristic polynomial is expressed by
	\[ \det ( \lambda  {\bf I}_{N } - {\bf U} )
	=( \lambda -1 )^{N - (s+t) } ( \lambda +1 )^{t-s} \prod_{j=t-s+1}^t ( \lambda-\alpha_j^{(+)})( \lambda-\alpha_j^{(-)}).  \]
Then we obtain 
$(N-s-t)$-multiple eigenvalue $1$, 
$(t-s)$-multiple eigenvalue $-1$ and 
$2s$ eigenvalues 
$\alpha_{q,j}^{(\pm)}$ $(j=t-s+1,\dots,t)$.
\item $N-(s+t)< 0$, $s-t\geq 0$ case: \\
Since $N-(s+t)< 0$, $(\lambda-1)^{N-(s+t)}$ is a negative power term. To cancel down it, 
$\{ ( \lambda-\alpha_j^{(+)}),\;( \lambda-\alpha_j^{(-)}) \}_{j=1}^t$ must contain $(s+t)-N$ terms of $(\lambda-1)$. 
Remark that if $\lambda=1$, then $\lambda_{q,j}=1$ from the above quadratic equation. So $\lambda_{q,N-s+1}=\cdots=\lambda_{q,t}=1$.
By the above consideration, the characteristic polynomial is expressed by
	\[ \det ( \lambda  {\bf I}_{N } - {\bf U} )
	=( \lambda -1 )^{(s+t)-N } ( \lambda +1 )^{s-t} \prod_{j=1}^{N-s} ( \lambda-\alpha_j^{(+)})( \lambda-\alpha_j^{(-)}).  \]
Then we obtain 
$(s+t-N)$-multiple eigenvalue $1$, 
$(s-t)$-multiple eigenvalue $-1$ and 
$2(N-s)$ eigenvalues 
$\alpha_{q,j}^{(\pm)}$ $(j=1,\dots,N-s)$.
\item $N-(s+t)< 0$, $s-t< 0$ case: \\
Since $N-(s+t)< 0$ and $s-t< 0$, both $(\lambda-1)^{N-(s+t)}$ and $(\lambda+1)^{s-t}$ are negative power terms. To cancel down it, 
$\{ ( \lambda-\alpha_j^{(+)}),\;( \lambda-\alpha_j^{(-)}) \}_{j=1}^t$ must contain $(s+t)-N$ terms of $(\lambda-1)$ and $t-s$ terms of $(\lambda+1)$. 
From the arguments of cases (2) and (3), we have $\lambda_{q,N-s+1} =\cdots=\lambda_{q,t}=1$ and $\lambda_{q,1}=\cdots=\lambda_{q,t-s}=0$. 
By the above consideration of  the characteristic polynomial is expressed by
	\[ \det ( \lambda  {\bf I}_{N } - {\bf U} )
	=( \lambda -1 )^{(s+t)-N } ( \lambda +1 )^{t-s} \prod_{j=t-s+1}^{N-s} ( \lambda-\alpha_j^{(+)})( \lambda-\alpha_j^{(-)}).  \]
Then we obtain 
$(s+t-N)$-multiple eigenvalue $1$, 
$(t-s)$-multiple eigenvalue $-1$ and 
$2(N-t)$ eigenvalues 
$\alpha_{q,j}^{(\pm)}$ $(j=t-s+1,\dots,N-s)$.
\end{enumerate}
Compiling the four cases, we have the desired conclusion. $\Box$

\section{The characteristic polynomial of the unitary matrix of a 2-tessellable staggered quantum matrix}

Let $G$ be a connected graph with $ \nu $ vertices and $ \epsilon $ edges, and let ${\bf U} = {\bf U}_1 {\bf U}_0 $ be 
the unitary matrix of a 2-tessellable SQW on $G$ such that both ${\bf U}_0 $ and ${\bf U}_1 $ are orthogonal reflections. 
Furthermore, let $\alpha $ and $ \beta $ be tessellations of ${\bf U} $ corresponding to ${\bf U}_0 $ and ${\bf U}_1 $, 
respectively. 
Set $\alpha = \{ \alpha {}_1 , \ldots , \alpha {}_m \} $ and $ \beta = \{ \beta {}_1 , \ldots , \beta {}_n \} $. 
Then we have 
\[
| \alpha {}_k \rangle = \sum_{ k^{\prime } \in \alpha {}_k } a_{k k^{\prime } } | k^{\prime } \rangle \  (1 \leq k \leq m), \  
| \beta {}_l \rangle = \sum_{ l^{\prime } \in \beta {}_l } b_{l l^{\prime } } | l^{\prime } \rangle \  (1 \leq l \leq n) , 
\]
\[
{\bf U}_0 =2 \sum^m_{k=1} | \alpha {}_k \rangle \langle \alpha {}_k | - {\bf I}_{ \nu} , \ 
{\bf U}_1 =2 \sum^n_{l=1} | \beta {}_l \rangle \langle \beta {}_l | - {\bf I}_{ \nu} . 
\]

Now, let $X$ be a finite  nonempty set and $S= \{ S_1 , \ldots , S_r \} $ a family of subsets of $X$. 
Then the {\em generalized intersection graph} $\Omega (S)$ is defined as follows: 
$V( \Omega (S))=S= \{ S_1 , \ldots , S_r \} $; $S_i $ and $S_j $ are joined by $|S_i \cap S_j |$ edges in 
$\Omega (S)$. 

Peterson \cite{Pet 2003} gave a necessary and sufficient condition for a graph to be 2-tessellable.

\begin{pro}[Perterson]\label{Peterson}
A graph $G$ is 2-tessellable if and only if $G$ is the line graph of a bipartite graph. 
\end{pro}

{\bf Sketch of proof}   Let $G$ be a 2-tessellable graph with two tessellations $\alpha $ and $\beta $. 
Set $S= \alpha \cup \beta $ and $H=\Omega (S)$. 
Then $H$ is a bipartite multi graph with partite set $\alpha $ and $\beta $. 
Furthermore, we have $G=L(\Omega (S))$. 

Conversely, it is clear that the line graph of a bipartite graph is 2-tessellable. 
Q.E.D.

By Proposition \ref{Peterson}, we can rewrite $| \alpha {}_k \rangle$ and $| \beta {}_l \rangle$. 
Let $H= \Omega ( \alpha \cup \beta )$ be a bipartite graph with bipartition $X= \{ x_1 , \ldots , x_m \} $, 
$Y= \{ y_1 , \ldots , y_n \} $ such that $G=L(H)$. 
Furthermore, we set $\alpha {}_k =N(x_k )(1 \leq k \leq m)$ and $\beta {}_l =N(y_l )(1 \leq l \leq n)$, 
where $N(x)= \{ e \in E(H) \mid x \in e \} $. 
Then we can write 
\[
| \alpha {}_k \rangle = \sum_{e \in N(x_k )} a_e |e \rangle \ (1 \leq k \leq m), \  
| \beta {}_l \rangle = \sum_{ f \in N(y_l )} b_f |f \rangle \  (1 \leq l \leq n) , 
\]
where $a_e ( or \ b_f )$ corresponds to $a_{k k^{\prime }}( or \ b_{l l^{\prime }})$ if 
$k^{\prime } ( or \ l^{\prime } ) \in V(G)$ corresponds to an edge $e ( or \ f) \in E(H)$. 

Now, we define an $m \times m$ matrix $\hat{{\bf A}} =( a_{x x^{\prime}} )_{x , x^{\prime } \in X} $ as follows: 
\[
a_{x x^{\prime}} :=\sum_{P=(e,f) : a \ (x, x^{\prime } )-path \ of \ length \ two \ in \ H} 
\overline{a}_{e} b_{e} a_{f} \overline{b}_{f} . 
\]

Then we obtain the following formula for the unitary matrix of a SQW on a 2-tessellable graph. 

\begin{thm}\label{staggeredQW}
Let $G$ be a connected 2-tessellable graph with $\nu$ vertices and $\epsilon $ edges, and let ${\bf U} = {\bf U}_1 {\bf U}_0 $ be 
the unitary matrix of a 2-tessellable SQW on $G$ such that both ${\bf U}_0 $ and ${\bf U}_1 $ are orthogonal reflections. 
Furthermore, let $\alpha $ and $ \beta $ be tessellations of ${\bf U} $ corresponding to ${\bf U}_0 $ and ${\bf U}_1 $, 
respectively. 
Set $| \alpha |=m$ and $| \beta |=n$. 
Then, for the unitary matrix ${\bf U} = {\bf U}_1 {\bf U}_0 $, we have    
\[
\det ( {\bf I}_{\nu } -u {\bf U} )=(1-u )^{ \nu -m-n} (1+u )^{n-m} \det ((1+u)^2 {\bf I}_m -4u \hat{{\bf A}} ) . 
\]
\end{thm}

{\em Proof }.  Let $\alpha = \{ \alpha {}_1 , \ldots , \alpha {}_m \} $ and $ \beta = \{ \beta {}_1 , \ldots , \beta {}_n \} $. 
Then we have 
\[
| \alpha {}_k \rangle = \sum_{ k^{\prime } \in \alpha {}_k } a_{k k^{\prime } } | k^{\prime } \rangle \  (1 \leq k \leq m), \  
| \beta {}_l \rangle = \sum_{ l^{\prime } \in \beta {}_l } b_{l l^{\prime } } | l^{\prime } \rangle \  (1 \leq l \leq n) , 
\]
\[
{\bf U}_0 =2 \sum^m_{k=1} | \alpha {}_k \rangle \langle \alpha {}_k | - {\bf I}_{ \nu} , \ 
{\bf U}_1 =2 \sum^n_{l=1} | \beta {}_l \rangle \langle \beta {}_l | - {\bf I}_{ \nu} . 
\]
Furthermore,  $H= \Omega ( \alpha \cup \beta )$ is expressed as follows: 
\[
V(H)=X \cup Y: \  a \ bipartition, \ X= \{ x_1 , \ldots , x_m \} , \ Y= \{ y_1 , \ldots , y_n \} ; 
\]
\[
N(x_k )= \{ e_{k1} , \ldots , e_{kd_k } \} , \ d_k = \deg {}_H x_k \ (1 \leq k \leq m) ;
\]
\[
N( y_l )= \{ f_{l1} , \ldots , e_{l \overline{d}_l } \} , \ \overline{d}_l = \deg {}_H y_l \ (1 \leq l \leq n) ,
\]
where $N(x)= \{ e \in E(H) \mid x \in e \} , \ x \in V(H)$ and $d_1 + \cdots + d_m = \overline{d}_1 + \cdots + \overline{d}_n = \nu$. 

We consider $\alpha {}_k =N(x_k )(1 \leq k \leq m)$ and $\beta {}_l =N(y_l )(1 \leq l \leq n)$. 
By Proposition \ref{Peterson}, we can write 
\[
| \alpha {}_k \rangle = \sum_{e \in N(x_k )} a_e |e \rangle \ (1 \leq k \leq m), \  
| \beta {}_l \rangle = \sum_{ f \in N(y_l )} b_f |f \rangle \  (1 \leq l \leq n) , 
\]
\[
{\bf U}_0 =2 \sum^m_{k=1} | \alpha {}_k \rangle \langle \alpha {}_k | - {\bf I}_{ \nu} , \ 
{\bf U}_1 =2 \sum^n_{l=1} | \beta {}_l \rangle \langle \beta {}_l | - {\bf I}_{ \nu} \  and \ {\bf U} = {\bf U}_1 {\bf U}_0 . 
\]

Now, let $x=x_1 \in X$, $d=d_x = \deg x$ and $N(x)= \{ e_1 , \ldots , e_d \}$. 
Set $\alpha {}_x = \alpha {}_i (x=x_i )$.  
Then  the submatrix of $| \alpha {}_x \rangle \langle \alpha {}_x | $ corresponding to the $e_1, \ldots , e_d $ rows 
and the $e_1, \ldots , e_d $ columns is we have 
\[
\left[ 
\begin{array}{cccc}
|a_{e_1} |^2 & a_{e_1} \overline{a}_{e_2} & \cdots & a_{e_1} \overline{a}_{e_d} \\
a_{e_2} \overline{a}_{e_1} & |a_{e_2} |^2 & \cdots & a_{e_2} \overline{a}_{e_d} \\
\vdots & \vdots & \ddots & \vdots \\
a_{e_d} \overline{a}_{e_1} & a_{e_d} \overline{a}_{e_2} & \cdots & | a_{e_d} |^2 
\end{array} 
\right]
,  
\] 
where $E(H)= \{ e_1 , \ldots , e_d , \ldots \} $. 
Thus, the submatrix of ${\bf U}_0  =2 \sum^m_{i=1} | \alpha {}_i \rangle \langle \alpha {}_i | - {\bf I}_{ \nu } $ 
corresponding to the $e_1, \ldots , e_d $ rows and the $e_1, \ldots , e_d $ columns is   
\[
\left[ 
\begin{array}{cccc}
2 |a_{e_1} |^2 -1 & 2 a_{e_1} \overline{a}_{e_2} & \cdots & 2 a_{e_1} \overline{a}_{e_d} \\
2 a_{e_2} \overline{a}_{e_1} & 2 |a_{e_2} |^2 -1 & \cdots & 2 a_{e_2} \overline{a}_{e_d} \\
\vdots & \vdots & \ddots & \vdots \\
2 a_{e_d} \overline{a}_{e_1} & 2 a_{e_d} \overline{a}_{e_2} & \cdots & 2| a_{e_d} |^2 -1 
\end{array} 
\right]
. 
\]

Let $y=y_1 $, $d^{\prime } =d_y = \deg y$ and $N(y)= \{ f_1 , \ldots , f_{d^{\prime }} \}$. 
Similarly to ${\bf U}_0 $, the submatrix of ${\bf U}_1  =2 \sum^n_{j=1} | \beta {}_j \rangle \langle \beta {}_j | - {\bf I}_{ \nu } $ 
corresponding to the $f_1 , \ldots , f_{d^{\prime }} $ rows and the $f_1 , \ldots , f_{d^{\prime }} $ columns is  we have 
\[
\left[ 
\begin{array}{cccc}
2 |b_{f_1} |^2 -1 & 2 b_{f_1} \overline{b}_{f_2} & \cdots & 2 b_{f_1} \overline{b}_{f_{d^{\prime }} } \\
2 b_{f_2} \overline{b}_{f_1} & 2 |b_{f_2} |^2 -1 & \cdots & 2 b_{f_2} \overline{b}_{f_{d^{\prime }} } \\
\vdots & \vdots & \ddots & \vdots \\
2 b_{f_{d^{\prime }} } \overline{b}_{f_1} & 2 b_{f_{d^{\prime }} } \overline{b}_{f_2} & \cdots & 2| b_{f_{d^{\prime } }} |^2 -1 
\end{array} 
\right]
. 
\]

Now, let ${\bf K} =( {\bf K}_{ex} )$ ${}_{e \in E(H); x \in X} $ be 
the $\nu \times m$ matrix defined as follows: 
\[
{\bf K}_{ex} :=\left\{
\begin{array}{ll}
a_e & \mbox{if $x \in e$, } \\
0 & \mbox{otherwise. } 
\end{array}
\right.
\]
Furthermore, we define the $\nu \times n$ matrix 
${\bf L} =( {\bf L}_{ey} )_{e \in E(H); y \in Y} $ as follows: 
\[
{\bf L}_{ey} :=\left\{
\begin{array}{ll}
b_e & \mbox{if $y \in e$, } \\
0 & \mbox{otherwise. } 
\end{array}
\right.
\] 
Then we have 
\[
{\bf K} \ {}^*{\bf K} = \sum^m_{k=1} | \alpha {}_k \rangle \langle \alpha {}_k | , \ 
{\bf L} \ {}^*{\bf L} = \sum^n_{l=1} | \beta {}_l \rangle \langle \beta {}_l | . 
\]
Furthermore, since 
\[
\sum_{e \in N(x)} |a_e |^2 = \sum_{f \in N(y)} |b_f |^2 =1 \  
for \  each \  x \in X \  and \  y \in Y ,  
\]
we have 
\[
{}^*{\bf K} {\bf K} = {\bf I}_m \  and \ {}^*{\bf L} {\bf L} = {\bf I}_n . 
\]

Therefore, by Theorem 3.1, it follows that 
\[
\det ( {\bf I}_{\nu } -u {\bf U} ) 
= (1-u )^{ \nu -m-n} (1+u)^{n-m}  
\det ((1+u )^2 {\bf I}_{m}-4u \ {}^*{\bf K} {\bf L} \ {}^*{\bf L} {\bf K} ) . 
\]

But, we have 
\[
( {}^*{\bf K} {\bf L} )_{xy} = \overline{a}_e b_e \  for \ e=xy \in E(G) . 
\]
Furthermore, we have 
\[
{}^*{\bf K} {\bf L} \ {}^*{\bf L} {\bf K} =( {}^*{\bf K} {\bf L} ) \ {}^* ( {}^*{\bf K} {\bf L} ) . 
\] 
Thus, for $x, x^{\prime } \in X$, 
\[
( {}^*{\bf K} {\bf L} \ {}^*{\bf L} {\bf K} )_{x x^{\prime }} 
= \sum_{P=(e,f) : a \ (x, x^{\prime } )-path \ of \ length \ two \ in \ H} 
\overline{a}_{e} b_{e} a_{f} \overline{b}_{f} . 
\]
Thus, we have 
\[
{}^*{\bf K} {\bf L} \ {}^*{\bf L} {\bf K} = \hat{{\bf A}} . 
\]
Hence, 
\[
\det ( {\bf I}_{\nu } -u {\bf U} )=(1-u )^{ \nu -m-n} (1+u )^{n-m} \det ((1+u)^2 {\bf I}_m -4u \hat{{\bf A}} ). 
\] 
$\Box$

By Theorem~\ref{staggeredQW} and Corollary~\ref{cor1}, we obtain the following. 
\begin{cor}
Let $G$ be a connected 2-tessellable graph with $\nu$ vertices and $\epsilon $ edges, and let ${\bf U} = {\bf U}_1 {\bf U}_0 $ be 
the unitary matrix of a 2-tessellable SQW on $G$ such that both ${\bf U}_0 $ and ${\bf U}_1 $ are orthogonal reflections. 
Furthermore, let $\alpha $ and $ \beta $ be tessellations of ${\bf U} $ corresponding to ${\bf U}_0 $ and ${\bf U}_1 $, 
respectively. 
Set $| \alpha |=m$ and $| \beta |=n$. 
Then, for the unitary matrix ${\bf U} = {\bf U}_1 {\bf U}_0 $, we have    
\[
\det ( \lambda  {\bf I}_{ \nu } - {\bf U} )=( \lambda -1 )^{ \nu -m-n} ( \lambda +1 )^{n-m} \det (( \lambda +1)^2 {\bf I}_m -4 \lambda \hat{{\bf A}} ). 
\]
\end{cor}
By Corollary~\ref{cor2p}, we obtain the spectrum of ${\bf U}$. 
\begin{cor}
Let $G$ be a connected 2-tessellable graph with $\nu$ vertices and $\epsilon $ edges, and let ${\bf U} = {\bf U}_1 {\bf U}_0 $ be 
the unitary matrix of a 2-tessellable SQW on $G$ such that both ${\bf U}_0 $ and ${\bf U}_1 $ are orthogonal reflections. 
Furthermore, let $\alpha $ and $ \beta $ be tessellations of ${\bf U} $ corresponding to ${\bf U}_0 $ and ${\bf U}_1 $, 
respectively. 
Set $| \alpha |=m$ and $| \beta |=n$. 
Then  the spectra of the unitary matrix ${\bf U} = {\bf U}_1 {\bf U}_0 $ are given as follows: 
Let $0\leq \lambda_{p,1}\leq \cdots \leq \lambda_{p,m}$ be the eigenvalues of $\hat{{\bf A}}$. 
\begin{enumerate} 
\item $2(\mathrm{Max}\{n,\nu-m\}-\mathrm{Max}(0,n-m))$ eigenvalues: 
\[
\lambda = e^{ \pm 2i \theta} , \ \cos {}^2 \theta \in \{\lambda_{p,j}\in Spec ( \hat{{\bf A}} ) \;\mid\; j=\mathrm{Max}\{1,m-n+1\},\dots, \mathrm{Max}\{m,\nu-n\} \},   
\]
\item $|\nu -m-n|$ eigenvalues: 1; 
\item $|n-m|$ eigenvalues: -1. 
\end{enumerate} 
\end{cor}

\section{The characteristic polynomial of the Szegedy matrix}

We present a formula for the characteristic polynomial of the Szegedy matrix of a bipartite graph. 
Let $G=(X \sqcup Y, E)$ be a connected multi-bipartite graph with partite set $X$ and $Y$. 
Moreover, set $|V(G)|= \nu $, $|E|=|E(G)|= \epsilon $, $|X|=m$ and $|Y|=n$. 
Then we consider the Hilbert space ${\cal H} = \ell {}^2 (E)=span \{  |e \rangle \mid e \in E \}$.
Let $p: E \rightarrow [0,1]$ and $q: E \rightarrow [0,1]$ be the functions such that 
\[
\sum_{X(e)=x} p(e)= \sum_{Y(e)=y} q(e)=1, \forall x\in X , \forall y \in Y , 
\]
where $X(e)$ and $Y(e)$ are the vertex of $e$ belonging to $X$ and $Y$, respectively. 

Let ${\bf W} = {\bf R}_1 {\bf R}_0 $ be a Szegedy matrix of $G$, where 
\[
{\bf R}_0  =2 \sum_{x \in X} | \phi {}_x \rangle \langle \phi {}_x | - {\bf I}_{ \epsilon } , \ 
{\bf R}_1  =2 \sum_{y \in Y} | \psi {}_y \rangle \langle \psi {}_y | - {\bf I}_{ \epsilon } , 
\]
\[
| \phi {}_x \rangle = \sum_{X(e)=x} \sqrt{p(e)} |e \rangle \  and \ 
| \psi {}_y \rangle = \sum_{Y(e)=y} \sqrt{q(e)} |e \rangle . 
\]

Then we define an $m \times m$ matrix ${\bf A}_p =( a^{(p)}_{x x^{\prime }} )_{x, x^{\prime } \in X} $ 
as follows: 
\[
a^{(p)}_{x x^{\prime}} :=\sum_{P=(e,f): a \ (x, x^{\prime } )-path \ of \ length \ two \ in \ G} 
\sqrt{p(e) q(e) p(f) q(f)} . 
\]
Note that 
\[
a^{(p)}_{xx} = \sum_{x \in e} p(e) q(e) , x \in X .  
\]

Then the characteristic polynomial of the Szegedy matrix of a bipartite graph is given as follows. 

\begin{thm}\label{SzegedyMat}
Let $G=(X \sqcup Y, E)$ and ${\bf W} $ be as the above. 
Then, for the Szegedy matrix ${\bf W} = {\bf R}_1 {\bf R}_0 $, we have    
\[
\det ( {\bf I}_{\epsilon } -u {\bf W} )=(1-u )^{ \epsilon - \nu} (1+u )^{n-m} \det ((1+u)^2 {\bf I}_m -4u {\bf A}_p ). 
\]
\end{thm}

{\em Proof }.  Let $X= \{ x_1 , \ldots , x_m \} $ and $Y= \{ y_1 , \ldots , y_n \}$. 
Let $x \in X$ and $y \in Y$. 
Then, let  
\[
| \phi {}_x \rangle = \sum_{X(e)=x} \sqrt{p(e)} |e \rangle \  and \ 
| \psi {}_y \rangle = \sum_{Y(e)=y} \sqrt{q(e)} |e \rangle . 
\] 

Now, let $x \in X$, $d=d_x = \deg x$ and $N(x)= \{ e_1 , \ldots , e_d \}$. 
Moreover, set $e_j =x y_{k_j} $ and $p_{xj} = p(x y_{k_j} )$ for $j=1, \ldots , d$. 
Then the submatrix of $| \phi {}_x \rangle \langle \phi {}_x | $ corresponding to the $e_1, \ldots , e_d $ rows 
and the $e_1, \ldots , e_d $ columns is  
\[
\left[ 
\begin{array}{cccc}
p_{x1} & \sqrt{ p_{x1} p_{x2} } & \cdots & \sqrt{ p_{x1} p_{xd} } \\
\sqrt{ p_{x2} p_{x1} } & p_{x2} & \cdots & \sqrt{ p_{x2} p_{xd} } \\
\vdots & \vdots & \ddots & \vdots \\
\sqrt{ p_{xd} p_{x1} } & \sqrt{ p_{xd} p_{x2} } & \cdots & p_{xd} 
\end{array} 
\right]
. 
\]
Thus, the submatrix of ${\bf R}_0  =2 \sum_{x \in X} | \phi {}_x \rangle \langle \phi {}_x | - {\bf I}_{ \epsilon } $ 
corresponding to the $e_1, \ldots , e_d $ rows and the $e_1, \ldots , e_d $ columns is   
\[
\left[ 
\begin{array}{cccc}
2 p_{x1} -1 & 2 \sqrt{ p_{x1} p_{x2} } & \cdots & 2 \sqrt{ p_{x1} p_{xd} } \\
2 \sqrt{ p_{x2} p_{x1} } & 2 p_{x2} -1 & \cdots & 2 \sqrt{ p_{x2} p_{xd} } \\
\vdots & \vdots & \ddots & \vdots \\
2 \sqrt{ p_{xd} p_{x1} } & 2 \sqrt{ p_{xd} p_{x2} } & \cdots & 2 p_{xd} -1 
\end{array} 
\right]
. 
\]

Let $y \in Y$, $d^{\prime } =d_y = \deg y$ and $N(y)= \{ f_1 , \ldots , f_{d^{\prime }} \}$. 
Moreover, set $f_j =y x_{k_l} $ and $q_{yl} = q(y x_{k_l} )$ for $l=1, \ldots , d^{\prime }$. 
Similarly to ${\bf R}_0 $, the submatrix of ${\bf R}_1 =2 \sum_{x \in X} | \psi {}_y \rangle \langle \psi {}_y | - {\bf I}_{ \epsilon } $ 
corresponding to the $f_1 , \ldots , f_{d^{\prime }} $ rows and the $f_1 , \ldots , f_{d^{\prime }} $ columns is  
\[
\left[ 
\begin{array}{cccc}
2 q_{y1} -1 & 2 \sqrt{ q_{y1} q_{y2} } & \cdots & 2 \sqrt{ q_{y1} p_{y d^{\prime } } } \\
2 \sqrt{ q_{y2} q_{y1} } & 2 q_{y2} -1 & \cdots & 2 \sqrt{ q_{y2} q_{y d^{\prime } } } \\
\vdots & \vdots & \ddots & \vdots \\
2 \sqrt{ q_{yd^{\prime } } q_{y1} } & 2 \sqrt{ q_{yd^{\prime } } q_{y2} } & \cdots & 2 q_{y d^{\prime } } -1 
\end{array} 
\right]
. 
\]

Now, let ${\bf K} =( {\bf K}_{ex} )$ ${}_{e \in E(G); x \in X} $ be 
the $\epsilon \times m$ matrix defined as follows: 
\[
{\bf K}_{ex} :=\left\{
\begin{array}{ll}
\sqrt{ p(e) } & \mbox{if $x \in e$, } \\
0 & \mbox{otherwise. } 
\end{array}
\right.
\]
Furthermore, we define the $\epsilon \times n$ matrix 
${\bf L} =( {\bf L}_{ey} )_{e \in E(G); y \in Y} $ as follows: 
\[
{\bf L}_{ey} :=\left\{
\begin{array}{ll}
\sqrt {q(e) } & \mbox{if $y \in e$, } \\
0 & \mbox{otherwise. } 
\end{array}
\right.
\] 
Moreover, since 
\[
\sum_{X(e)=x} p(e)= \sum_{Y(e)=y} q(e)=1, \forall x \in X , \forall y \in Y , 
\]
we have 
\[
{}^t {\bf K} {\bf K} = {\bf I}_m \  and \  {}^t {\bf L} {\bf L} = {\bf I}_n . 
\]

Thus, by Theorem 3.1, for ${\bf W} = {\bf R}_1 {\bf R}_0 $ and $|u|<1$, 
\[
\det ( {\bf I}_{\epsilon } -u {\bf W} ) 
	=  (1-u )^{\epsilon - \nu } (1+u)^{n-m}  
		\det ((1+u )^2 {\bf I}_{m}-4u \ {}^t {\bf K} {\bf L} \ {}^t {\bf L} {\bf K} ) . 
\]

But, we have 
\[
( {}^t {\bf K} {\bf L} )_{xy} = \sum_{X(e)=x, Y(e)=y} \sqrt{ p(e) q(e) } .  
\]
Furthermore, we have 
\[
{}^t {\bf K} {\bf L} \ {}^t {\bf L} {\bf K} =( {}^t {\bf K} {\bf L} ) \ {}^t ( {}^t {\bf K} {\bf L} ) . 
\] 
Thus, for $x, x^{\prime } \in X (x \neq x^{\prime })$, 
\[
( {}^t {\bf K} {\bf L} \ {}^t {\bf L} {\bf K} )_{x x^{\prime }} 
= \sum_{P=(e,f): a \ (x, x^{\prime } )-path \ of \ length \ two \ in \ G} \sqrt{p(e) q(e) p(f) q(f)} . 
\]
In the case of $x= x^{\prime }$,  
\[
( {}^t {\bf K} {\bf L} \ {}^t {\bf L} {\bf K} )_{xx} 
= \sum_{X(e)=x} p_e q_e . 
\]
Therefore, it follows that 
\[
{}^t {\bf K} {\bf L} \ {}^t {\bf L} {\bf K} = {\bf A}_p . 
\]
Hence, 
\[
\det ( {\bf I}_{\epsilon } -u {\bf W} )=(1-u )^{\epsilon - \nu } (1+u )^{n-m} \det ((1+u)^2 {\bf I}_m -4u {\bf A}_p ). 
\] 
$\Box$

By Theorem~\ref{SzegedyMat} and Corollary~\ref{cor1}, we obtain the following. 
\begin{cor}
Let $G=(X \sqcup Y, E)$ and ${\bf W} $ be as the above. 
Then, for the Szegedy matrix ${\bf W} = {\bf R}_1 {\bf R}_0 $, we have    
\[
\det ( \lambda  {\bf I}_{\epsilon } - {\bf W} )=( \lambda -1 )^{\epsilon - \nu} ( \lambda +1 )^{n-m} \det (( \lambda +1)^2 {\bf I}_m -4 \lambda {\bf A}_p ). 
\]
\end{cor}
By Theorem~\ref{SzegedyMat} and Corollary~\ref{cor2q}, we obtain the spectrum of ${\bf W}$, which is consistency with \cite{Sze}.  
\begin{cor}
Let $G=(X \sqcup Y, E)$ and ${\bf W} $ be as the above. 
Suppose that $n \geq m$. 
Then, the spectra of the Szegedy matrix ${\bf W} = {\bf R}_1 {\bf R}_0 $ are given as follows: \\
If $G$ is not a tree, then 
\begin{enumerate} 
\item $2m$ eigenvalues: 
\[
\lambda = e^{\pm 2i\theta}, \ \cos^2 \theta\in Spec ( {\bf A}_p ) ; 
\]
\item $\epsilon - \nu$ eigenvalues: 1; 
\item $n-m$ eigenvalues: -1. 
\end{enumerate} 

If $G$ is a tree, then 
\begin{enumerate} 
\item $2m-2$ eigenvalues: 
\[
\lambda = e^{\pm 2i\theta}, \ \cos^2 \theta \in Spec ( {\bf A}_p\setminus\{1\} ) ; 
\]
\item one eigenvalue: 1; 
\item $n-m$ eigenvalues: -1. 
\end{enumerate} 
\end{cor}

Similarly, if $n<m$, then the following result holds. 

\begin{cor}
Let $G=(X \sqcup Y, E)$ and ${\bf W} $ be as the above. 
Suppose that $m \geq n$. 
Then we define an $n \times n$ matrix ${\bf A}_q =( a^{(q)}_{y y^{\prime }} )_{y, y^{\prime } \in Y} $ 
as follows: 
\[
a^{(q)}_{y y^{\prime}} :=\sum_{Q=(e,f): a \ (y, y^{\prime } )-path \ of \ length \ two \ in \ G} 
\sqrt{p(e) q(e) p(f) q(f)} . 
\]
Note that 
\[
a^{(q)}_{xx} = \sum_{y \in e} p(e) q(e) , y \in Y .  
\]

Then, the spectra of the Szegedy matrix ${\bf W} = {\bf R}_1 {\bf R}_0 $ are given as follows: 
If $G$ is not a tree, then 
\begin{enumerate} 
\item $2n$ eigenvalues: 
\[
\lambda = e^{\pm 2i\theta}, \ \cos^2 \theta \in Spec ( {\bf A}_q ) ; 
\]
\item $\epsilon - \nu$ eigenvalues: 1; 
\item $m-n$ eigenvalues: -1. 
\end{enumerate} 

If $G$ is a tree, then 
\begin{enumerate} 
\item $2n-2$ eigenvalues: 
\[
\lambda = e^{\pm 2i\theta}, \ \cos^2 \theta \in Spec ( {\bf A}_q\setminus\{ 1\} ) ; 
\]
\item one eigenvalue: 1; 
\item $m-n$ eigenvalues: -1. 
\end{enumerate} 
\end{cor}

\section{An example} 

Let $G=K_{2,2} $ be the complete bipartite graph with partite set $X= \{ a,b \} , Y= \{ c,d \} $. 
Then we arrange edges of $G$ as follows: 
\[
e_1 =ac, \ e_2 = ad, \ e_3 =bc, \ e_4 =bd . 
\] 
Furthermore, we consider the following two functions $p: E \rightarrow [0,1]$ and $q: E \rightarrow [0,1]$ 
such that 
\[
p(e_1 )=p(e_2 )=p(e_3 )=p(e_4 )=1/2 \ and \ q(e_1 )=q(e_2 )=q(e_3 )=q(e_4 )=1/2 . 
\]

Now, we have 
\[
| \phi {}_a \rangle =
\left[ 
\begin{array}{c}
1/ \sqrt{2} \\
1/ \sqrt{2} \\
0 \\ 
0
\end{array} 
\right]
, 
| \phi {}_b \rangle =
\left[ 
\begin{array}{c}
0 \\
0 \\
1/ \sqrt{2} \\
1/ \sqrt{2} 
\end{array} 
\right]
, 
| \psi {}_c \rangle =
\left[ 
\begin{array}{c}
1/ \sqrt{2} \\
0 \\
1/ \sqrt{2} \\
0 
\end{array} 
\right]
, 
| \psi {}_d \rangle =
\left[ 
\begin{array}{c}
0 \\
1/ \sqrt{2} \\
0 \\ 
1/ \sqrt{2} 
\end{array} 
\right]
. 
\]
Thus, we have 
\[
{\bf K} = 
\left[ 
\begin{array}{cc}
1/ \sqrt{2} & 0 \\
1/ \sqrt{2} & 0 \\
0 & 1/ \sqrt{2} \\
0 & 1/ \sqrt{2} 
\end{array} 
\right]
, 
{\bf L} = 
\left[ 
\begin{array}{cc}
1/ \sqrt{2} & 0 \\
0 & 1/ \sqrt{2} \\
1/ \sqrt{2} & 0 \\
0 & 1/ \sqrt{2} 
\end{array} 
\right] 
.
\]
Therefore, it follows that 
\[
{\bf K} \ {}^t {\bf K} = 
\left[ 
\begin{array}{cccc}
1/2 & 1/2 & 0 & 0 \\
1/2 & 1/2 & 0 & 0 \\
0 & 0 & 1/2 & 1/2 \\
0 & 0 & 1/2 & 1/2 
\end{array} 
\right]
, 
\]
\[
{\bf L} \ {}^t {\bf L} = 
\left[ 
\begin{array}{cccc}
1/2 & 0 & 1/2 & 0 \\
0 & 1/2 & 0 & 1/2 \\
1/2 & 0 & 1/2 & 0 \\
0 & 1/2 & 0 & 1/2
\end{array} 
\right]
. 
\]
Hence, 
\[
{\bf R}_0  =2 \sum_{x \in X} | \phi {}_x \rangle \langle \phi {}_x | - {\bf I}_{4} =
\left[ 
\begin{array}{cc}
{\bf J}_0 & {\bf 0} \\
{\bf 0} & {\bf J}_0  
\end{array} 
\right] 
, 
\]
\[
{\bf R}_1  =2 \sum_{y \in Y} | \psi {}_y \rangle \langle \psi {}_y | - {\bf I}_{4}  =
\left[ 
\begin{array}{cc}
{\bf 0} & {\bf I}_2 \\
{\bf I}_2 & {\bf 0} 
\end{array} 
\right] 
,  
\]
where 
\[
{\bf J}_0  =
\left[ 
\begin{array}{cc}
0 & 1\\
1 & 0 
\end{array} 
\right] 
. 
\] 
Thus, 
\[
{\bf W} = {\bf R}_1 {\bf R}_0 = 
\left[ 
\begin{array}{cc}
{\bf 0} & {\bf J}_0 \\  
{\bf J}_0 & {\bf 0}   
\end{array} 
\right] 
. 
\]

But, 
\[
{\bf A}_p ={}^t {\bf K} {\bf L} \ {}^t {\bf L} {\bf K} = 
\left[ 
\begin{array}{cc}
1/2 & 1/2 \\  
1/2 & 1/2   
\end{array} 
\right] 
. 
\]
Thus, 
\[
\det ( \lambda {\bf I}_2 - {\bf A}_p )= ( \lambda -1/2)^2 -1/4= \lambda ( \lambda -1) . 
\]
Therefore, it follows that 
\[
Spec ( {\bf A}_p )= \{ 1, 0 \} . 
\]
Furthermore, since $m=n=2$, we have $mn-m-n=n-m=0$. 
By Corollary 4.3, the eigenvalues of ${\bf W}$ are 
\[
\lambda =1, 1, -1, -1 .    
\]
There are eigenvalues induced from ${\bf A}_p $. 

\section{The characteristic polynomial of the modified time evolution matrix of the duplication of the modified digraph}

Let $G$ be a connected graph with $n$ vertices and $\epsilon$ edges which may have multiple edges and self loops 
, and the duplication graph be $G_2$.
We set $p,q: E(G_2)\to [0,1]$ so that $\sum_{V(e)=v}p(e)=\sum_{V'(e)=v'}q(e)=1$ with 
\[\{p(e) \;|\; e\in E_{G_2}(v,u')\}=\{q(f) \;|\; f\in E_{G_2}(v',u) \} \]
for any $v\in V$ and $u'\in V'$. Thus $q$ is determined by $p$. 
The $2n \times 2n$ stochastic matrix ${\bf P}$ is denoted by 
	\[ ({\bf P})_{u,v}=p_{uv}
        	= \begin{cases} \sum_{V(e)=u,\;V'(e)=v} p(e) & \text{if $u\in V$, $v\in V'$, } \\ 
                		\sum_{V'(e)=u,\;V(e)=v} q(e) & \text{if $u\in V'$, $v\in V$, } \\
                                0 & \text{otherwise.}
                \end{cases} \]
	
Furthermore, let $M$ be a set of $m$ marked vertices in $G$, 
and the modified bipartite graph of the duplication graph $G_2$ with the marked element 
	\[ M_2 =M \cup M' \; (M' = \{v' \;|\; v \in M \}) \]  
be denoted by $\vec{G}_2$. 
Let ${\bf W}^{\prime } = {\bf R}^{\prime }_1 {\bf R}^{\prime }_0 $ be the modified time evolution matrix of 
the modified Szegedy walk on $\ell^2(\tilde{E}_2)$. 
Here $E_M=E(G_2)\cup [N_2]$, where $[N_2]$ is set of the matching edges between marked elements and its copies, 
that is $[N_2]=\{ mm' \;|\; m\in M \}$. 
Thus the cardinality of $\ell^2(E_M)$ is $2\epsilon+m$. 
Under the setting of ${\bf W'}$, we took the modification of $p$ and $q$ as follows. 
Let $p',q': E_M \to [0,1]$ be 
\[
p^{\prime } (e) :=\left\{
\begin{array}{ll}
p(e) & \mbox{if $V(e)\notin M$, } \\
1 & \mbox{if $e\in [N_2]$,} \\
0 & \mbox{otherwise,} 
\end{array}
\right.
\]
\[
q^{\prime } (f) :=\left\{
\begin{array}{ll}
q(f) & \mbox{if $V'(f) \not\in M^{\prime }$,} \\
1 & \mbox{if $f\in [N_2]$, } \\
0 & \mbox{otherwise, }
\end{array}
\right.
\]
where
\[
\sum_{V(e)=x} p^{\prime } (e)= \sum_{V'(e)=y} q^{\prime } (e)=1, \forall x \in V , \forall y \in V' .  
\]  
The modified $2n \times 2n$ stochastic matrix ${\bf P}'$ is given by changing $p$ and $q$ to $p'$ and $q'$ as follows: 
		\[ ({\bf P}^{\prime } )_{u,v}=p'_{uv}
        	= \begin{cases} \sum_{V(e)=u,\;V'(e)=v} p'(e) & \text{if $u\in V$, $v\in V'$, } \\ 
                		\sum_{V'(e)=u,\;V(e)=v} q'(e) & \text{if $u\in V'$, $v\in V$, } \\
                                0 & \text{otherwise.}
                \end{cases} \]

The reflection operators ${\bf R}_0'$ and ${\bf R}_1'$ are described by $\{\phi'_v\}_{v\in V}$ and $\{\psi'_u\}_{u\in V'}$ as follows: 
	\begin{align*} 
        {\bf R}_0' &= 2\sum_{v\in V}|\phi'_v\rangle\langle \phi'_v|-{\bf I}_{2\epsilon+m}, \\
        {\bf R}_1' &= 2\sum_{u\in V'}|\psi'_u\rangle\langle \psi'_u|-{\bf I}_{2\epsilon+m}, 
        \end{align*}
where $\phi'_v=\sum_{V(e)=v}\sqrt{p'(e)}|e\rangle$, $\psi'_v=\sum_{V^{\prime } (e)=v}\sqrt{q'(e)}|e\rangle$. 
See Sect. 2.5 for more detailed this setting. 
Let $\{|v\rangle\}_{v\in V}$ be the standard basis of $\mathbb{C}^{n}$, that is, 
$(|v\rangle)_{u}=1$ if $v=u$, $(|v\rangle)_{u}=0$ otherwise, 
where $n=|V|$. 
We define $(2\epsilon+m)\times n$ matrices as follows, where $2\epsilon=|E(G_2)|$: 
	\begin{align*}
        K &= \sum_{v\in V(e)} |\phi'_v \rangle\langle v|, \\
        L &= \sum_{u'\in V'(e)} |\psi'_{u'} \rangle\langle u| ,  
        \end{align*}
that is, 
\[
K_{ev} :=
\left\{
\begin{array}{ll}
\sqrt{ p^{\prime } (e) } & \mbox{if $V(e)=v$, } \\
0 & \mbox{otherwise,  } 
\end{array}
\right.
\]
\[
L_{ev} :=
\left\{
\begin{array}{ll}
\sqrt{ q^{\prime } (e)} & \mbox{if $V'(e)=v'$, } \\
0 & \mbox{otherwise. } 
\end{array}
\right.
\]
Let $r$ be the number of edges connecting non-marked elements and its copies, that is, 
	\[ r=|\{ e\in E_M \;|\; V(e)\notin M, \; V'(e)\notin M'\}|. \]
Let $s$ be the number of edges connecting non-marked elements and copies of marked elements, that is, 
	\[ s=|\{ e\in E_M \;|\; V(e)\notin M, \; V'(e)\in M'\}|. \]
We set $\epsilon {}^{\prime } = r+2s+m$. 
Remark that if there is no marked element connecting to another marked element in the original graph $G$, 
then $\epsilon' =2 \epsilon+m$, on the other hand, if not, $\epsilon' <2 \epsilon+m$ since such an edge 
connecting marked element in $G$ is omitted in the procedure making $\vec{G}_2$ from $G$. 
By the definitions of ${\bf R}^{\prime }_0 $ and ${\bf R}^{\prime }_1 $, 
${\bf K} \ {}^t {\bf K} $ is equal to $\sum_{x \in X} | \phi {}^{\prime }_x \rangle \langle \phi {}^{\prime }_x |$, 
and ${\bf L} \ {}^t {\bf L} $ is equal to $\sum_{y \in Y} | \psi {}^{\prime }_y \rangle \langle \psi {}^{\prime }_y |$. 
Thus,   
\[
{\bf R}^{\prime }_0 =2 \sum_{x \in X} | \phi {}^{\prime }_x \rangle \langle \phi {}^{\prime }_x | - {\bf I}_{2\epsilon+m } =      
2 {\bf K} \ {}^t {\bf K} - {\bf I}_{2\epsilon+m  } , 
\]
\[
{\bf R}^{\prime }_1 =2 \sum_{y \in Y} | \psi {}^{\prime }_y \rangle \langle \psi {}^{\prime }_y | - {\bf I}_{2\epsilon+m  } =      
2 {\bf L} \ {}^t {\bf L} - {\bf I}_{2\epsilon+m } .   
\]

Now, we define an $n \times n$ matrix $\hat{{\bf A}}^{\prime }_p $ as follows: 
\[
\hat{{\bf A}}^{\prime }_p = {}^t {\bf K} {\bf L} \ {}^t {\bf L} {\bf K} ,  
\]
Remark that $q$ is determined by $p$ and so as $p'$ and $q'$. 
The $v,u$ element of this symmetric matrix $\hat{{\bf A}}^{\prime }_p$ is computed as follows: 
${}^t{\bf LK}$ is expressed by 
	\begin{align*} 
        {}^t{\bf LK} 
        &= \begin{bmatrix} \langle \psi'_{v_1'}| \\ \vdots \\ \langle \psi'_{v_n'}| \end{bmatrix} 
           \begin{bmatrix} |\phi'_{v_1}\rangle & \cdots & |\phi'_{v_n}\rangle \end{bmatrix} \\
        \\
        & =\begin{bmatrix}
        \langle \psi'_{v'_1} | \phi'_{v_1} \rangle & \cdots & \langle \psi'_{v'_1} | \phi'_{v_n}\rangle \\
        \vdots & \ddots & \vdots \\
        \langle \psi'_{v'_n} | \phi'_{v_1} \rangle & \cdots & \langle \psi'_{v'_n} | \phi'_{v_n} \rangle
        \end{bmatrix}
        \end{align*}
Thus 
	\begin{align*} 
        ({}^t{\bf LK})_{u,v}
        	&=\langle \psi'_{u} | \phi'_{v}\rangle 
        	= \sum_{e\in E_M} \overline{\psi'_{u}(e)} \phi'_{v}(e) \\
                &= \sum_{V(e)=v,\;V'(e)=u} \sqrt{p'(e)q'(e)}, 
        \end{align*}
which is the summation of a real valued weight over all the path from $u\in V'$ to $v\in V$. 
Therefore
	\[ (\hat{{\bf A}}^{\prime }_p)_{u,v} = \sum_{(e,f): (u,v)-\mathrm{path\;in\;}G_2} \sqrt{p'(e)q'(e)p'(f)q'(f)}. \]
Since $p'(e)=0$, $q'(f)=0$ for every ``$V(e)\in M$, $V'(e)\notin M$" and ``$V'(f)\in M'$, $V(f)\notin M$", 
\[(\hat{{\bf A}}^{\prime }_p)_{u,v} = 
	\begin{cases}
        \sum_{(e,f)\in Q_2} \sqrt{p(e)q(e)p(f)q(f)} & \text{ if $u,v\in V\setminus M$, }\\
        \delta_{u,v} & \text{ if $u,v\in M$,} \\
        0 & \text{ otherwise. }
        \end{cases}
        \]
Here the summation $Q_2$ is over all the $2$-length path in $G_2$ from $u\in V$ to $v\in V$ never going into $M$ and $M'$. 

If the following condition holds, we say $p$ satisfies the detailed balanced condition: 
there exists $\pi: V \sqcup V'$ such that 
	\[ p'(e)\pi(V(e))=q'(e)\pi(V'(e)) \]
for every $e\in E_M$ with $V(e)\notin M$ and $V'(e)\notin M'$, and $\pi(u)=1$ if $u\in M \sqcup M'$. 
A typical setting of $p(e)=1/\mathrm{deg}(V(e))$ and $q(e)=1/\mathrm{deg}(V'(e))$ satisfies the detailed balanced condition 
by $\pi(u)=\mathrm{deg}(u)$ for every $u\in (V\setminus M)\sqcup(V'\setminus M')$. 
If the detailed balanced condition holds, 
Since the values $q(e)$ and $p(f)$, where $(e,f)$ is $(u,v)$-path in $G_2$, are equivalent to 
	\[ q(e)=\frac{\pi (V(e))}{\pi(V'(e))}=\frac{\pi(u)}{\pi(V'(e))},\;p(f)=\frac{\pi (V'(f))}{\pi(V(f))}=\frac{\pi'(v)}{\pi(V'(e))},  \]
we have
	\[ \sqrt{p(e)q(e)p(f)q(f)}=\sqrt{\pi(u)/\pi(v)} p(e)q(f).  \]
Then it is expressed by 
	\[ (\hat{{\bf A}}^{\prime }_p)_{u,v}
        	= \begin{cases}
                \sqrt{\pi(u)/\pi(v)} \sum_{(e,f)\in Q_2} p(e)q(f) & \text{ if $u,v\notin M$, } \\
                \delta_{u,v} & \text{ if $u,v\in M$ } \\
                0 & \text{otherwise, } 
                \end{cases}
        \]
Therefore if the detailed balanced condition holds, $\hat{{\bf A}}^{\prime }_p$ 
is unitary equivalent to the square of ${\bf P}_M':={\bf P}_M \oplus {\bf I}_{m}$,
where ${\bf P}_M$ is an $(n-m)\times (n-m)$ matrix describing the random walk  with the Dirichlet boundary condition at $M$: 
for $u,v\notin M$, 
	\[ ({\bf P}_M)_{u,v}
        	=\sum_{e\in E(G_2) \mathrm{\;with\;} V(e)=u,V'(e)=v'} p(e). \]
Thus 
	\[ ({\bf P}_M')_{u,v}
        	=\begin{cases}
                ({\bf P}_M)_{u,v} & \text{if $u,v\notin M$, } \\
                \delta_{u,v} & \text{if $u,v\in M$,} \\
                0 & \text{otherwise. }
                \end{cases} \]
Now we are in the place to give
the following formula for the the modified time evolution matrix of 
the modified Szegedy walk on
$\ell^2(E_M)$.  
\begin{thm}\label{QuantumSerch}
Let $G$ be a connected graph with $n$ vertices and $\epsilon$ edges 
which may have multiple edges and self loops. 
Let ${\bf W}^{\prime } = {\bf R}^{\prime }_1 {\bf R}^{\prime }_0 $ be the modified time evolution matrix of 
the modified Szegedy walk on 
$\ell^2(E_M)$ induced by random walk $p:E(G_2)\to [0,1]$ and the marked element $M$ with $|M|=m$. 

Then, for ${\bf W}^{\prime } $, we have    
\[
\det ( {\bf I}_{2\epsilon+m} -u {\bf W}^{\prime } )
=
(1-u )^{ 2(\epsilon-n)+m } 
\det ((1+u)^2 {\bf I}_n -4u \hat{{\bf A}}^{\prime }_p ). 
\]
In particular, if $p$ satisfies the detailed balanced condition, then 
\begin{align*}
\det ( {\bf I}_{2\epsilon+m} -u {\bf W}^{\prime } )
&=
(1-u )^{ 2(\epsilon-n)+3m } \det ((1+u)^2 {\bf I}_{n-m} -4u {\bf P}_M^2 ). 
\end{align*}
\end{thm}

{\em Proof }.  
The subset of edges connecting marked elements and its copies in $E_M$  denotes $F_M$, that is, 
	\[ F_M=\{ e\in E_M \;|\; V(e)\in M,\;V'(e)\in M' \}. \]
The cardinality of $F_M=2\epsilon+m-\epsilon'$. 
The definitions of $p'$ and $q'$ give $p'(e)=q'(e)=0$ for $e\in F_M$, 
which implies $\langle e|\phi'_v\rangle=\langle e|\psi'_u\rangle=0$ for any $u,v\in V$. 
Thus 
\begin{align*}
({\bf K} \; {}^t{\bf K})_{e,f} &= \sum_{v\in V} \langle e|\phi'_v\rangle \langle \phi'_v|f\rangle = 0, \\
({\bf L} \; {}^t{\bf L})_{e,f} &= \sum_{v\in V} \langle e|\psi'_v\rangle \langle \psi'_v|f\rangle = 0
\end{align*}
for every $e,f\in F_M$. 
Concerning the above, it holds that 
\begin{align*}
{\bf R}'_0 &=  2{\bf K}\; {}^t{\bf K}- {\bf I}_{2\epsilon+m} \\
	   &= (2{\bf K}\; {}^t{\bf K}- {\bf I}_{\epsilon'}) \oplus (-{\bf I}_{2\epsilon+m-\epsilon'}) \\
{\bf R}'_1 &= (2{\bf L}\; {}^t{\bf L}- {\bf I}_{2\epsilon+m}) \\
	   &= (2{\bf L}\; {}^t{\bf L}- {\bf I}_{\epsilon'}) \oplus (-{\bf I}_{2\epsilon+m-\epsilon'})
\end{align*}
Therefore if $F_M\neq \emptyset$, then 
\[ {\bf W}^{\prime } = (2{\bf L}\;{}^t{\bf L}- {\bf I}_{\epsilon'})(2{\bf K}\;{\bf {}^tK}- {\bf I}_{\epsilon'}) \oplus {\bf I}_{2\epsilon+m-\epsilon'}. \]
Therefore, if $F_M\neq \emptyset$, then at least $|F_M|$-multiple eigenvalue $1$ of ${\bf W}'$ exists. 

From now on we consider the second term of the above RHS. 
To this end, it is not a loss of generality that we take the assumption that $F_M=\emptyset$ 
putting $2{\bf K}\;{\bf {}^tK}- {\bf I}_{\epsilon'}={\bf R}'_0$, $2{\bf L}\;{\bf {}^tL}- {\bf I}_{\epsilon'}={\bf R}'_1$ and 
${\bf W}'={\bf R}'_1{\bf R}'_0$. 
Since 
\[
\sum_{V(e)=x} p^{\prime } (e)= \sum_{V^{\prime } (e)=y} q^{\prime } (e)=1, \forall x \in X , \forall y \in Y , 
\]
we have 
\[
{}^t {\bf K} {\bf K} = {}^t {\bf L} {\bf L} = {\bf I}_n . 
\]

Therefore, by Theorem 3.1, it follows that 
\[
\det ( {\bf I}_{\epsilon {}^{\prime } } -u {\bf W}^{\prime } ) \\ 
= (1-u )^{ \epsilon {}^{\prime } -2n}   
\det ((1+u )^2 {\bf I}_{n}-4u \ {}^t {\bf K} {\bf L} \ {}^t {\bf L} {\bf K} ) . 
\]

But, 
\[
\hat{{\bf A}}^{\prime }_p = {}^t {\bf K} {\bf L} \ {}^t {\bf L} {\bf K} . 
\]
Hence, 
If $F_M=\emptyset$, then
\[
\det ( {\bf I}_{\epsilon {}^{\prime } } -u {\bf W}^{\prime } )=(1-u )^{ \epsilon {}^{\prime } -2n} \det ((1+u)^2 {\bf I}_n -4u \hat{{\bf A}}^{\prime }_p ). 
\] 
Therefore if $F_M\neq \emptyset$, then 
\begin{align*}
\det ( {\bf I}_{2 \epsilon+m} - u {\bf W}^{\prime } )
	& =(1-u)^{2\epsilon+m-\epsilon'}\times (1-u )^{ \epsilon {}^{\prime} -2n} \det ((1+u)^2 {\bf I}_n -4u \hat{{\bf A}}^{\prime }_p ) \\
        &= (1-u)^{2(\epsilon-n)+m} \det ((1+u)^2 {\bf I}_n -4u \hat{{\bf A}}^{\prime }_p )
\end{align*}
Concerning the fact that $F_M=\emptyset$ if and only if $\epsilon'=2\epsilon+m$, then we have obtained the desired conclusion. 
If the detailed balanced condition holds, $\hat{{\bf A}}^{\prime }_p= ({\bf D}\oplus {\bf I}_{m}){{\bf P}_M'}^2 ({\bf D}^{-1}\oplus {\bf I}_{m})$, 
${\bf D}$ is an $(n-m)\times (n-m)$ diagonal matrix $\mathrm{diag}[\sqrt{\pi(u)}\;|\; u \notin M]$, that is, 
$({\bf D}\oplus {\bf I}_{m})|u\rangle = \sqrt{\pi(u)}$ if $u\notin M$, $({\bf D}\oplus {\bf I}_{m})|u\rangle = |u\rangle$ if $u\in M$.  
$\Box$

By Theorem~\ref{QuantumSerch} and Corollary~\ref{cor1}, we have obtain following. 
\begin{cor}\label{vovo}
Let $G$ be a connected graph with $n$ vertices and $\epsilon$ edges 
which may have multiple edges and self loops. 
Let ${\bf W}^{\prime } = {\bf R}^{\prime }_1 {\bf R}^{\prime }_0 $ be the modified time evolution matrix of 
the modified Szegedy walk on 
$\ell^2(E_M)$ induced by random walk $p:E(G_2)\to [0,1]$ and the marked element $M$ with $|M|=m$. 
Then, for the ${\bf W}^{\prime } = {\bf R}^{\prime }_1 {\bf R}^{\prime }_0$, we have    
\[
\det ( \lambda  {\bf I}_{ 2\epsilon+m}  - {\bf W}^{\prime } )
=( \lambda -1 )^{ 2(\epsilon-n)+m}  \det (( \lambda +1)^2 {\bf I}_n -4 \lambda \hat{{\bf A}}^{\prime }_p ). 
\]
\end{cor}

By Theorem~\ref{QuantumSerch} and Corollary~\ref{cor2q}, we obtain the eigenvalues of ${\bf W}'$. 
\begin{cor}
Let $G$ be a connected graph with $n$ vertices and $\epsilon$ edges 
which may have multiple edges and self loops. 
Let ${\bf W}^{\prime } = {\bf R}^{\prime }_1 {\bf R}^{\prime }_0 $ be the modified time evolution matrix of 
the modified Szegedy walk on 
$\ell^2(E_M)$ induced by random walk $p:E(G_2)\to [0,1]$ and the marked element $M$ with $|M|=m$. 
Then  the spectra of the unitary matrix ${\bf W}^{\prime } = {\bf R}^{\prime }_1 {\bf R}^{\prime }_0$ are given as follows: 
\begin{enumerate} 
\item If $2(\epsilon-n)+m\geq 0$, that is, ``$G$ is not a tree" or ``$m>1$", then 
\begin{enumerate}
\item $2n$ eigenvalues: 
\[
\lambda = e^{\pm 2i\theta}, \ \cos^2\theta \in Spec ( \hat{{\bf A}}^{\prime }_p ) ; 
\]
\item $2(\epsilon-n) + m $ eigenvalues: 1. 
\end{enumerate}
\item  
otherwise, that is, $G$ is a tree and $m\in \{0,1\}$, then 
\begin{enumerate}
\item $2(n-1)$ eigenvalues: 
\[
\lambda = e^{\pm 2i\theta}, \ \cos^2\theta \in Spec ( \hat{{\bf A}}^{\prime }_p )\setminus{\{1\}}; 
\]
\item $m$-multiple eigenvalue $1$. 
\end{enumerate}
\end{enumerate}
\end{cor}
{\em Proof }.  Since $\epsilon-n<0$ if and only if $G$ is a tree, thus $2(\epsilon-n)+m<0$ if and only if $G$ is a tree and $m\in \{0,1\}$. 
By Corollary~\ref{vovo}, 
\[ \det ( \lambda  {\bf I}_{ 2\epsilon+m } - {\bf W}^{\prime } )
=( \lambda -1 )^{ 2(\epsilon-n)+m } \prod_{j=1}^{n} (\lambda-\alpha_j^{(+)})(\lambda-\alpha_j^{(-)})  \]
holds, where $\alpha_j^{(\pm)}$ are the solutions of $\lambda^2-2(2\mu-1)\lambda+1=0$ with $\nu\in Spec(\hat{{\bf A}}^{\prime }_p)$. 
The second term has $2n=2\epsilon+2$ solutions while the dimension of the total space is now $2\epsilon+m$. 
But in this situation since $2(\epsilon-n)+m=-2+m <0$, then the power of the first term $(1-\lambda)^{2(\epsilon-n)+m}$ is negative.
Thus the second term should includes the $(\lambda-1)^{(2-m)}$ term counteracted by the first term. 
The result follows. 
$\Box$

\section{An example} 

Let $G=K_{3} $ be the complete graph with three vertices $v_1 , v_2 , v_3 $, and 
${\bf P} =( p_{uv} )_{u,v \in V(G)} $ the following stochastic matrix of $G$: 
\[
{\bf P} = \frac{1}{2} 
\left[ 
\begin{array}{ccc}
0 & 1 & 1 \\
1 & 0 & 1 \\ 
1 & 1 & 0 
\end{array} 
\right]
. 
\]
Furthermore, let $M= \{ v_3 \} $ be a set of $m$ marked vertices in $G$. 
%
Thus we set $E_M$ by 
\[ \{e_1,e_2,f_1,f_2,f_1',f_2',g\} \]
where $e_1=\{v_1,v_2'\}$, $e_2=\{v_2,v_1'\}$, $f_1=\{v_1,v_3'\}$, $f_2=\{v_2,v_3'\}$, $f_1'=\{v_3,v_2'\}$, $f_2'=\{v_3,v_1'\}$ and $g=\{v_3,v_3'\}$. 
The duplication graph of $G$ is denoted by $G_2$. $E_M$ is the union of $E(G_2)$ and $\{ g \} $. 
The modified stochastic matrix ${\bf P}^{\prime } =( p^{\prime }_{uv} )_{u,v \in V(G_2)} $ derived from ${\bf P}$ with the marked element $M=\{v_3\}$ is 
given as follows: 
\[
{\bf P}^{\prime } =  
\left[ 
\begin{array}{cccccc}
0&0&0& 0 & 1/2 & 1/2 \\
0&0&0& 1/2 & 0 & 1/2 \\ 
0&0&0& 0 & 0 & 1 \\ 
0 & 1/2 & 1/2 & 0&0&0\\
1/2 & 0 & 1/2 & 0&0&0\\ 
0 & 0 & 1 & 0&0&0\\ 
\end{array} 
\right] 
\]
which means 
\begin{align*} 
	p'(e_1) &=p'(e_2)=p'(f_1)=p'(f_2)=1/2, \; p'(f_1')=p'(f_2')=0, \; p'(g)=1  \mathrm{\;and} \\
   	q'(e_1) &=q'(e_2)=q'(f_1')=q'(f_2')=1/2, \; q'(f_1)=q'(f_2)=0, \; q'(g)=1 
\end{align*}
 
Then the dimension of the total state space is 
\[
|E_M|=2\epsilon+m=\epsilon {}^{\prime } =2+2 \cdot 2+1=7 . 
\]
We put $X=\{v_1,v_2,v_3\}$ and its copy $X'=\{v_1',v_2',v_3'\}$. 
The $7 \times 3$ matrix $K$ is an incidence matrix between $7$ edges  $e_1 , e_2 , f_1 , f_2 , f^{\prime }_1 , f^{\prime }_2 , g$ 
and $X$ as follows: 
\[
{\bf K} =  
\left[ 
\begin{array}{ccc}
1/ \sqrt{2} & 0 & 0 \\
0 & 1/ \sqrt{2} & 0 \\
1/ \sqrt{2} & 0 & 0 \\
0 & 1/ \sqrt{2} & 0 \\
0 & 0 & 0 \\
0 & 0 & 0 \\
0 & 0 & 1 
\end{array} 
\right]
. 
\]
Furthermore, the $7 \times 3$ matrix $L$ is an incidence matrix between $7$ edges 
$e_1 , e_2 , f_1 , f_2 , f_1 , f_2 , g$
and $Y$ as follows: 
\[
{\bf L} =  
\left[ 
\begin{array}{ccc}
0 & 1/ \sqrt{2} & 0 \\
1/ \sqrt{2} & 0 & 0 \\
0 & 0 & 0 \\
0 & 0 & 0 \\
0 & 0 & 0 \\
1/ \sqrt{2} & 0 & 0 \\
0 & 1/ \sqrt{2} & 0 \\
0 & 0 & 1 
\end{array} 
\right]
. 
\]
Thus, we have 
\[
{\bf K} \ {}^t {\bf K} = \sum_{x \in X} | \phi {}_x \rangle \langle \phi {}_x |= 
\left[ 
\begin{array}{ccccccc}
1/2 & 0 & 1/2 & 0 &  &  & 0 \\
0 & 1/2 & 0 & 1/2 &  &  &  \\
1/2 & 0 & 1/2 & 0 &  &  &  \\
0 & 1/2 & 0 & 1/2 &  &  &  \\ 
  &  &  &  &  0 &  &  \\
  &  &  &  &   & 0 &  \\
0 &  &  &  &  &  & 1 
\end{array} 
\right]
\]
and 
\[
{\bf L} \ {}^t {\bf L} = \sum_{y \in Y} | \psi {}_y \rangle \langle \psi {}_y |= 
\left[ 
\begin{array}{ccccccc}
1/2 & 0 & 0 & 0 & 0 & 1/2 & 0 \\ 
0 & 1/2 & 0 & 0 & 1/2 & 0 & 0 \\ 
0 & 0 & 0 & 0 & 0 & 0 & 0 \\
0 & 0 & 0 & 0 & 0 & 0 & 0 \\ 
0 & 1/2 & 0 & 0 & 1/2 & 0 & 0 \\
1/2 & 0 & 0 & 0 & 0 & 1/2 & 0 \\
0 & 0 & 0 & 0 & 0 & 0 & 1 
\end{array} 
\right]
. 
\]
Therefore, it follows that 
\[
{\bf R}^{\prime}_0 = 2 {\bf K} \ {}^t {\bf K} - {\bf I}_7 
=
\left[ 
\begin{array}{cccccc}
0 & 0 & 1 & 0 &  &  0 \\
0 & 0 & 0 & 1 &  &    \\
1 & 0 & 0 & 0 &  &    \\
0 & 1 & 0 & 0 &  &    \\ 
  &  &  &  &  - {\bf I}_2 &   \\
  &  &  &  &  & 1  \\ 
\end{array} 
\right]
\]
and 
\[
{\bf R}^{\prime}_1 = 2 {\bf L} \ {}^t {\bf L} - {\bf I}_7 
=
\left[ 
\begin{array}{ccccccc}
0 & 0 & 0 & 0 & 0 & 1 & 0 \\
0 & 0 & 0 & 0 & 1 & 0 & 0 \\
0 & 0 & -1 & 0 & 0 & 0 & 0 \\
0 & 0 & 0 & -1 & 0 & 0 & 0 \\
0 & 1 & 0 & 0 & 0 & 0 & 0  \\
1 & 0 & 0 & 0 & 0 & 0 & 0  \\
0 & 0 & 0 & 0 & 0 & 0 & 1 
\end{array} 
\right]
\]
Hence, 
\[
{\bf W}^{\prime} = {\bf R}^{\prime}_1 {\bf R}^{\prime}_0 = 
\left[ 
\begin{array}{ccccccc}
0 & 0 & 0 & 0 & 0 & 1 & 0 \\
0 & 0 & 0 & 0 & 1 & 0 & 0 \\
-1 & 0 & -1 & 0 & 0 & 0 & 0 \\
0 & -1 & 0 & 0 & 0 & 0 & 0 \\
0 & 0 & 0 & 1 & 0 & 0 & 0 \\ 
0 & 0 & 1 & 0 & 0 & 0 & 0 \\
0 & 0 & 0 & 0 & 0 & 0 & 1   
\end{array} 
\right]
. 
\]

Now, we have 
\[
{}^t {\bf K} {\bf L} = 
\left[ 
\begin{array}{ccc}
0 & 1/2 & 0 \\  
1/2 & 0 & 0 \\ 
0 & 0 & 1    
\end{array} 
\right] 
. 
\]
Thus, we have 
\[
\hat{{\bf A}}^{\prime }_p = {}^t {\bf K} {\bf L} \ {}^t {\bf L} {\bf K} = 
\left[ 
\begin{array}{ccc}
1/4 & 0 & 0 \\  
0 & 1/4 & 0 \\ 
0 & 0 & 1   
\end{array} 
\right] 
. 
\]
Thus, 
\[
\det ( \lambda {\bf I}_2 - \hat{{\bf A}}^{\prime }_p )=( \lambda -1)( \lambda -1/4)^2 . 
\]
Therefore, it follows that 
\[
Spec (  \hat{{\bf A}}^{\prime }_p )= \{ 1, 1/4 \} . 
\]
Furthermore, since $n=3$, we have $\epsilon {}^{\prime } -2n=7-6=1$. 
By Corollary 6.3, the eigenvalues of ${\bf W}^{\prime } $ are 
\[
\lambda = 1, 1 , 1, \frac{-1 \pm i \sqrt{3}}{2} , \frac{-1 \pm i \sqrt{3}}{2} .    
\]

\section*{Acknowledgments}
The first author is partially supported by the Grant-in-Aid for Scientific Research (Chal-
lenging Exploratory Research) of Japan Society for the Promotion of Science (Grant No.
15K13443). 
The second author is partially supported by the Grant-in-Aid for Scientific Research (C) of Japan
Society for the Promotion of Science (Grant No. 15K04985). 
The third author is partially supported by 
the Grant-in-Aid for Young Scientists (B) of Japan Society for the
Promotion of Science (Grant No. 25800088).

\begin{small}
\bibliographystyle{jplain}

\end{small}

\end{document}